\documentclass[useAMS,usenatbib,A4]{mn2e}

\usepackage{graphicx}  
\usepackage{ amssymb }
\usepackage{journals}
\usepackage[dvips]{color}
\newcommand{\teff}{$T_{\rm eff}$} 
\newcommand{\logg}{$\log g$} 
\newcommand{\kms}{km s$^{-1}$}
\newcommand{\vt}{$\xi_t$} 
\newcommand{\fei}{Fe\,{\sc i}}
\newcommand{\feii}{Fe\,{\sc ii}}

\newcommand\simgt{\lower.3ex\hbox{\gtsima}}

\newcommand{\strom}{Str{\" o}mgren}

\newcommand\msun{M$_{\odot}$}

\def\msun{$M_{\odot}$}

\usepackage{placeins}
\usepackage[modulo]{lineno}

\title[CNO in NGC 1851 and NGC 6752]{CNO abundances in the globular clusters
NGC 1851 and NGC 6752\thanks{Based on observations collected at the European
Southern Observatory, Chile (ESO Programmes 65.L-0165 and 084.D-0693).}} 
\author[D.\ Yong et al.]
{David Yong,$^{1}$\thanks{E-mail: david.yong@anu.edu.au}\thanks{Stromlo Fellow}  
Frank Grundahl$^2$ and 
John E.\ Norris$^1$. \\ 
$^{1}$Research School of Astronomy and Astrophysics, Australian
National University, Canberra, ACT 2611, Australia\\ 
$^{2}$Stellar Astrophysics Centre, Department of Physics and 
Astronomy, Aarhus University, Ny Munkegade 120, DK-8000 Aarhus C, Denmark\\
}
\begin{document}


\pagerange{\pageref{firstpage}$-$\pageref{lastpage}} \pubyear{2013}

\maketitle

\label{firstpage}

\begin{abstract}
We measure the C+N+O abundance sum in red giant stars in two Galactic globular
clusters, NGC 1851 and NGC 6752. NGC 1851 has a split subgiant branch which
could be due to different ages or C+N+O content while NGC 6752 is
representative of the least complex globular clusters. For NGC 1851 and NGC
6752, we obtain average values of A(C+N+O) = 8.16 $\pm$ 0.10 ($\sigma$ = 0.34)
and 7.62 $\pm$ 0.02 ($\sigma$ = 0.06), respectively. When taking into account
the measurement errors, we find a constant C+N+O abundance sum in NGC 6752. The
C+N+O abundance dispersion is only 0.06 dex, and such a result requires that
the source of the light element abundance variations does not increase the
C+N+O sum in this cluster.  For NGC 1851, we confirm a large spread in C+N+O.
In this cluster, the anomalous RGB has a higher C+N+O content than the
canonical RGB by a factor of four ($\sim$0.6 dex). This result lends further
support to the idea that the two subgiant branches in NGC 1851 are roughly
coeval, but with different CNO abundances. 
\end{abstract}


\section{INTRODUCTION}

Galactic globular clusters continue to pose a series of intriguing questions
concerning stellar evolution, stellar nucleosynthesis and chemical evolution.
First, it has been known for several decades that globular clusters exhibit
star-to-star variations in the CN and CH line strengths
\citep[e.g.,][]{smith87}. These molecular line strength variations are driven
by star-to-star abundance variations for the light elements from C to Al (see
reviews by \citealt{kraft94} and \citealt{gratton04,gratton12} for details.)
Secondly, a star-to-star dispersion in iron-peak elements, and other elements,
has long been known to exist in the globular cluster $\omega$ Centauri (e.g.,
\citealt{freeman75,cohen81,norris95,smith00,johnson10}). More recently,
abundance dispersions have also been identified in a number of globular
clusters including M2 \citep{yong14}, M22 \citep{marino09,marino11,roederer11},
M54 \citep{carretta10}, NGC 1851 \citep{yong081851,carretta11}, NGC
3201\footnote{Other studies of this cluster do not find evidence for an iron
dispersion \citep{Ca09,munoz13}.} \citep{simmerer13}, NGC 5824\footnote{This
result is based on metallicities from the calcium triplet.} \citep{dacosta14}
and Terzan 5 \citep{ferraro09,origlia13}, although the shape of the metallicity
distribution function differs between these objects. 

The light element abundance variations are believed to result from hydrogen
burning at high temperature \citep{denisenkov90,langer93,prantzos07}. The
astrophysical site in which these nuclear reactions occur continues to be
debated with asymptotic giant branch (AGB) stars, fast rotating massive stars,
massive binaries and supermassive stars among the candidates
\citep{fenner04,ventura05,karakas06,decressin07a,demink09,marcolini09,denissenkov14}.
Additionally, many details regarding the production of these abundance
variations including the initial mass function, minimum timescale, required
mass budget, degree of (or need for) dilution with pristine gas and star
formation modes still need to be established \citep{bastian13,renzini13}. An
important constraint on the site and nature of the nucleosynthesis comes from
the C+N+O\footnote{Here and throughout the paper, ``C+N+O'' is the sum of the
C, N and O abundances and these values are on the $\log\epsilon$ scale. For
example, the \citet{asplund09} solar values are $\log\epsilon$(C) = 8.43,
$\log\epsilon$(N) = 7.83 and $\log\epsilon$(O) = 8.69 and this gives C+N+O =
8.92.} abundance sum. In fast rotating massive stars, the C+N+O abundance sum
is expected to remain constant; the slow winds are enriched in H-burning
products whereas the He-burning products are ejected at later times at high
velocity \citep{decressin07a,decressin07b}. AGB stars, on the other hand, are
expected to increase the C+N+O abundance sum \citep{fenner04}. That said,
adjustments to the input physics can result in AGB models that produce an
essentially constant C+N+O abundance sum \citep{ventura05}. 

The dispersion in heavy element abundances in $\omega$ Cen has led to the
suggestion that it is the nucleus of an accreted dwarf galaxy
\citep{freeman93,bekki03}. For the recently discovered globular clusters with
dispersions in iron-peak elements, the chemical similarities with $\omega$
Centauri may also require a similarly complex formation process. The sequence
of events leading to the formation of these globular clusters remain poorly
understood. Some of these objects exhibit multiple subgiant branches, and it is
well known that the C+N+O abundance sum plays a key role in age determinations
based on subgiant branch analyses \citep{rood85}. In the case of NGC 1851, the
double subgiant branch \citep{milone08} could be composed of two coeval
populations with different mixtures of C+N+O abundances \citep{cassisi08}.
Since the discovery of the double subgiant branch in NGC 1851, understanding
its nature and formation history has been an active area of research
\citep{dantona09,han09,lee09,milone09,olszewski09,ventura09,zoccali09,carretta11,carretta12,bekki12,gratton121851,gratton121851b,lardo121851,joo13,marino14}.
Indeed, it has been suggested that NGC 1851 may be the product of the merger of
two clusters \citep{carretta101851}. 

Despite the importance of the C+N+O content in globular clusters, there is only
a modest number of studies on this topic (e.g.,
\citealt{brown91,dickens91,ivans99,cohen05,smith05,yong086712,marino11,marino12b}).
For NGC 1851, there are conflicting results regarding the C+N+O abundance sum
\citep{yong09,villanova10}. In the case of NGC 6752, \citet{carretta05} found
that the C+N+O abundance sum was constant, within a factor of $\sim$2, for
their sample of dwarfs and subgiants. We note that the C, N and O abundances
are typically derived from the CH molecular lines, CN molecular lines and [OI]
atomic lines, respectively. Due to molecular equilibrium, deriving the C
abundance requires knowledge of the O abundance, and vice versa. Similarly, N
abundances derived from the CN lines require knowledge of both the C and O
abundances. For N measurements from CN molecular lines, the uncertainties can
be magnified by the errors associated with both the C and O abundances. The
situation can be improved by analysis of the NH molecular lines. One advantage
is that the inferred N abundance requires no knowledge of the C and/or O
abundances. A clear disadvantage is that the best NH molecular lines are near
3360\,\AA; this is a crowded spectral region and red giant branch stars have
limited flux in the blue relative to redder wavelengths. 

The purpose of this paper is to examine the C+N+O abundance sum in the globular
clusters NGC 1851 and NGC 6752. NGC 6752 is representative of the least complex
globular clusters; it has a single subgiant branch (when not viewed using
filters sensitive to molecular lines of CH, CN, CO, NH and OH:
\citealt{milone13}) and no large dispersion in iron-peak elements (modulo the
small but statistically significant variations identified by \citealt{yong13}).
Measurements of C+N+O in this cluster would serve to constrain the origin of
the light element abundance variations in globular clusters and provide an
important baseline for comparison with multiple subgiant branch clusters. NGC
1851 is representative of multiple subgiant branch globular clusters with
star-to-star abundance dispersions for iron-peak and neutron-capture elements.
Measurements of C+N+O in this cluster would help establish whether or not the
double subgiant branch populations are coeval. The outline of the paper is as
follows. Section 2 describes the sample selection and observations. The
analysis is presented in Section 3. Section 4 includes the results and
discussion and we present concluding remarks in Section 5.

\section{SAMPLE SELECTION AND OBSERVATIONS}

\begin{figure*}
\centering
      \includegraphics[width=.7\hsize]{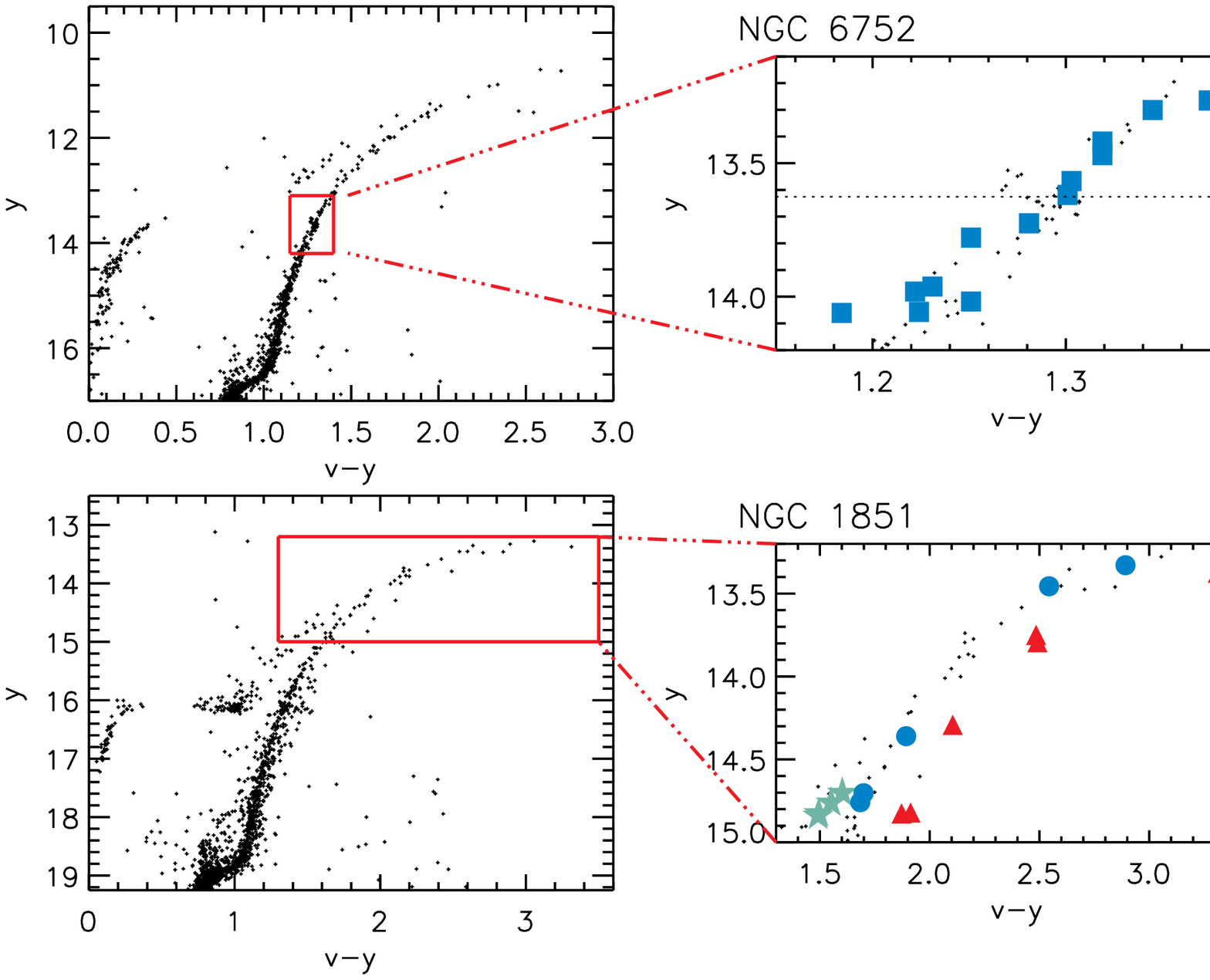}
      \caption{Colour-magnitude diagrams for $y$ versus $v-y$ for NGC 6752
(upper) and NGC 1851 (lower) using photometry from \citet{grundahl99}. The
right hand panels show a smaller region of the CMD. For NGC 6752, the program
stars are marked by large blue squares and the location of the RGB bump is
indicated by the dotted line. For NGC 1851, the aqua star symbols, blue circles
and red triangles refer to AGB, canonical RGB and anomalous RGB stars,
respectively. 
      \label{fig:cmd} }
\end{figure*}

The targets were selected from the \strom\ $uvby$ photometry from
\citet{grundahl99}, see Figure \ref{fig:cmd} and Tables \ref{tab:6752} and
\ref{tab:1851}. For NGC 6752, the targets were a subset of those observed by
\citet{grundahl02} and lie near the RGB bump (see Table \ref{tab:6752}). For
NGC 1851, the targets lie on the canonical RGB, the AGB and the anomalous
RGB\footnote{In the various tables, the anomalous RGB objects are denoted by
``m1'' since they were first noticed as unusually red in the CMDs involving the
$m1$ index in the \citet{grundahl99} photometry.} (see Table \ref{tab:1851}).
To avoid contamination from nearby stars, the targets were selected to have no
neighbours within 2.5\arcsec\ and $\Delta V$mag = 2.5. Based on their location
in CMDs, stellar parameters and radial velocities, all stars are likely cluster
members. 

\begin{table*}
 \centering
 \begin{minipage}{180mm}
  \caption{Stellar parameters and CNO abundances for NGC 6752.} 
  \label{tab:6752} 
  \begin{tabular}{@{}llcccccccccrcc@{}}
  \hline
	Name 1 &
	Name 2\footnote{Star names from \citet{buonanno86}.} &
	RA 2000 & 
	Dec.\ 2000 & 
	$V$ & 
	\teff\ &
	\logg\ &
	\vt\ &
	[Fe/H] &
	C &
	N$_{\rm NH}$ &
	N$_{\rm CN}$ &
	O &
	C+N+O\footnote{N is from NH.} \\
        (1) & 
        (2) &
        (3) &
        (4) & 
        (5) & 
        (6) & 
        (7) & 
        (8) & 
        (9) & 
        (10) & 
        (11) & 
        (12) & 
        (13) & 
        (14) \\ 
  \hline
NGC6752-1  & B2882  & 19 10 47 & $-$60 00 43 & 13.27 & 4749 & 1.95 & 1.41 & $-$1.58 & 6.41 & 6.36 & $<$7.36 & 7.60 & 7.65 \\ 
NGC6752-2  & B1635  & 19 11 11 & $-$60 00 17 & 13.30 & 4779 & 2.00 & 1.39 & $-$1.59 & 6.06 & 7.55 &    7.95 & 6.93 & 7.65 \\ 
NGC6752-4  & B611   & 19 11 33 & $-$60 00 02 & 13.42 & 4806 & 2.04 & 1.40 & $-$1.61 & 6.16 & 7.35 &    7.85 & 6.96 & 7.52 \\ 
NGC6752-6  & B3490  & 19 10 34 & $-$59 59 55 & 13.47 & 4804 & 2.06 & 1.40 & $-$1.61 & 6.36 & 7.30 &    7.60 & 7.10 & 7.54 \\ 
NGC6752-8  & B3103  & 19 10 45 & $-$59 58 18 & 13.56 & 4910 & 2.15 & 1.33 & $-$1.62 & 6.51 & 6.25 & $<$7.25 & 7.61 & 7.66 \\ 
NGC6752-9  & B3880  & 19 10 26 & $-$59 59 05 & 13.57 & 4824 & 2.11 & 1.38 & $-$1.63 & 6.61 & 6.05 & $<$7.05 & 7.64 & 7.69 \\ 
NGC6752-11 & B2728  & 19 10 50 & $-$60 02 25 & 13.62 & 4829 & 2.13 & 1.32 & $-$1.64 & 6.41 & 7.15 &    7.65 & 7.36 & 7.60 \\ 
NGC6752-15 & B2782  & 19 10 49 & $-$60 01 55 & 13.73 & 4850 & 2.19 & 1.35 & $-$1.61 & 6.56 & 5.70 & $<$7.20 & 7.70 & 7.73 \\ 
NGC6752-16 & B4446  & 19 10 15 & $-$59 59 14 & 13.78 & 4906 & 2.24 & 1.32 & $-$1.60 & 6.41 & 7.40 &    7.70 & 7.08 & 7.60 \\ 
NGC6752-19 & B1113  & 19 11 23 & $-$59 59 40 & 13.96 & 4928 & 2.32 & 1.29 & $-$1.61 & 6.56 & 6.90 &    7.45 & 7.32 & 7.51 \\ 
NGC6752-20 & \ldots & 19 10 36 & $-$59 56 08 & 13.98 & 4929 & 2.33 & 1.32 & $-$1.59 & 6.15 & 7.45 &    7.90 & 7.12 & 7.63 \\ 
NGC6752-21 & \ldots & 19 11 13 & $-$60 02 30 & 14.02 & 4904 & 2.33 & 1.29 & $-$1.61 & 6.51 & 6.95 &    7.45 & 7.51 & 7.65 \\ 
NGC6752-23 & B1668  & 19 11 12 & $-$59 58 29 & 14.06 & 4916 & 2.35 & 1.27 & $-$1.62 & 6.16 & 7.45 &    7.95 & 7.15 & 7.64 \\ 
NGC6752-24 & \ldots & 19 10 44 & $-$59 59 41 & 14.06 & 4948 & 2.37 & 1.15 & $-$1.65 & 6.51 & 5.95 & $<$7.45 & 7.53 & 7.58 \\ 
  \hline
\end{tabular}
\end{minipage}
\end{table*}

\begin{table*}
 \centering
 \begin{minipage}{180mm}
  \caption{Stellar parameters and CNO abundances for NGC 1851.} 
  \label{tab:1851} 
  \begin{tabular}{@{}llccccccccccccc@{}}
  \hline
	Name 1 & 
	Name 2\footnote{Star names from \citet{stetson81}.} & 
	RA 2000 & 
	Dec.\ 2000 & 
	$V$ & 
	CMD & 
	\teff\ &
	\logg\ &
	\vt\ &
	[Fe/H] &
	C &
	N\footnote{The N abundances are from CN and have been adjusted by
$-$0.44 dex (see Section 4.1 and 4.2 for details).} &
	O &
	C+N+O \\
        (1) & 
        (2) &
        (3) &
        (4) & 
        (5) & 
        (6) & 
        (7) & 
        (8) & 
        (9) & 
        (10) & 
        (11) & 
        (12) & 
        (13) &
        (14) \\ 
  \hline
NR  712 & 236    & 05 13 59.45 & $-$40 05 22.59 & 14.70 & RGB & 4392 & 1.42 & 1.50 & $-$1.33 & 6.26 &    6.76 &    7.73 &    7.79 \\ 
NR 1290 & 168    & 05 14 19.34 & $-$40 04 23.85 & 13.33 & RGB & 3738 & 0.27 & 1.95 & $-$1.26 & 6.26 &    7.96 &    7.83 &    8.21 \\ 
NR 4740 & 126    & 05 14 17.24 & $-$40 02 08.01 & 14.36 & RGB & 4259 & 1.19 & 1.50 & $-$1.20 & 6.31 &    7.11 &    7.73 &    7.84 \\ 
NR 6221 & \ldots & 05 14 07.78 & $-$40 01 18.15 & 14.76 & RGB & 4426 & 1.46 & 1.40 & $-$1.27 & 6.11 &    7.75 &    7.33 &    7.63 \\ 
NR 6250 & \ldots & 05 14 02.80 & $-$40 01 22.78 & 13.46 & RGB & 3924 & 0.54 & 1.85 & $-$1.30 & 6.16 &    7.81 &    7.68 &    8.06 \\ 
          \\
NR 1469 & 210    & 05 14 10.35 & $-$40 04 23.57 & 14.76 & AGB & 4639 & 1.54 & 1.80 & $-$1.32 & 6.21 &    8.16 &  \ldots &  \ldots \\ 
NR 2352 & \ldots & 05 14 13.52 & $-$40 03 40.88 & 14.84 & AGB & 4688 & 1.60 & 1.40 & $-$1.16 & 6.16 &    8.06 &  \ldots &  \ldots \\ 
NR 3272 & 137    & 05 14 15.02 & $-$40 03 04.08 & 14.83 & AGB & 4607 & 1.54 & 1.65 & $-$1.33 & 6.11 &    7.66 &    7.93 &    8.12 \\ 
NR 8066 & 22     & 05 14 10.35 & $-$39 58 14.83 & 14.70 & AGB & 4596 & 1.49 & 1.70 & $-$1.32 & 6.16 &    8.11 &    7.53 &    8.22 \\ 
          \\
NR 2953 & \ldots & 05 14 05.86 & $-$40 03 24.57 & 13.75 &  m1 & 3958 & 0.70 & 1.95 & $-$1.21 & 5.86 &    8.81 &    7.13 &    8.82 \\ 
NR 3213 & 112    & 05 14 25.96 & $-$40 02 53.78 & 13.79 &  m1 & 4002 & 0.76 & 2.20 & $-$1.30 & 6.21 &    8.46 &    7.33 &    8.49 \\ 
NR 5171 & \ldots & 05 14 06.60 & $-$40 02 02.63 & 14.83 &  m1 & 4315 & 1.42 & 1.55 & $-$1.34 & 6.26 &    8.16 &    7.28 &    8.22 \\ 
NR 5246 & \ldots & 05 14 01.33 & $-$40 02 05.81 & 13.37 &  m1 & 3666 & 0.18 & \multicolumn{6}{c}{Spectrum affected by TiO}  \\ 
NR 5543 & 319    & 05 13 50.30 & $-$40 02 06.98 & 14.82 &  m1 & 4402 & 1.47 & 1.70 & $-$1.30 & 6.46 &    8.21 &  \ldots &  \ldots \\ 
NR 6217 & 58     & 05 14 14.47 & $-$40 01 10.93 & 14.29 &  m1 & 4212 & 1.13 & 1.70 & $-$1.26 & 6.06 &    8.36 &    7.28 &    8.40 \\ 
  \hline
\end{tabular}
\end{minipage}
\end{table*}

\begin{figure*}
\centering
      \includegraphics[width=.7\hsize]{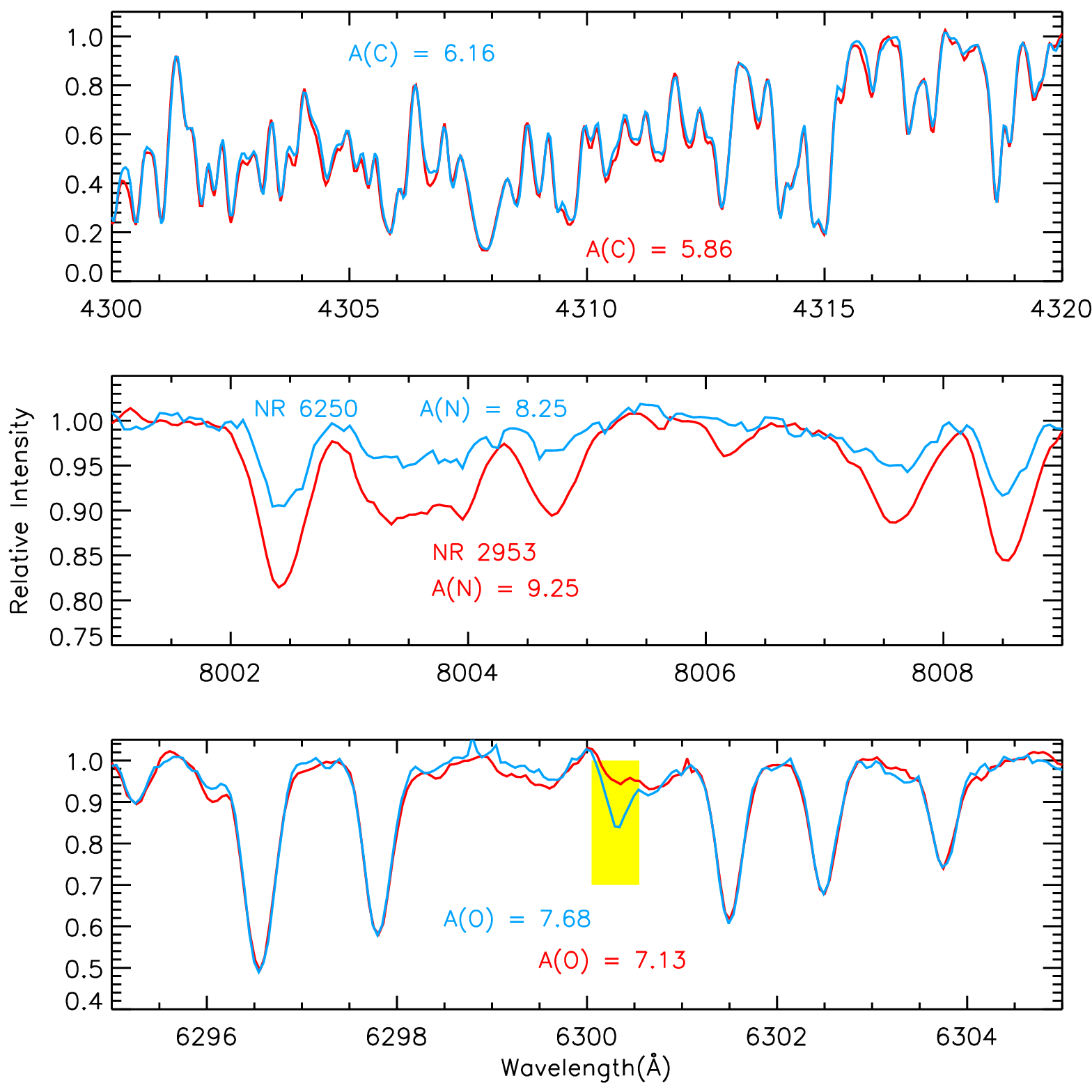}
      \caption{A portion of the spectra for two stars in NGC 1851 with similar
stellar parameters, but belonging to the two different RGBs. NR 6250 (blue) is
a canonical RGB star while NR 2953 (red) is an anomalous RGB star. The upper
and middle panels include regions used in determining the C and N abundances,
respectively. In the lower panel, the 6300\AA\ [OI] line is highlighted. The C,
N and O abundances for both stars are included in the panels. 
      \label{fig:1851spec} }
\end{figure*}

Observations of these targets were obtained using the multiobject spectrograph
FLAMES/GIRAFFE \citep{pasquini02} in IFU mode. The field of view is 25 arcmin
and there are 15 IFU units each of which has an aperture of 2$\arcsec$ $\times$
3$\arcsec$ consisting of 20 square spaxels of length 0.52$\arcsec$.  For NGC
1851, we obtained spectra using the HR4 ($R$ = 32,500 @ $\sim$4300\,\AA; total
exposure time of 7.9 hours), HR13 ($R$ = 36,000 @ $\sim$6300\,\AA, total
exposure time of 1.4 hours) and HR19B ($R$ = 35,000 @ $\sim$8000\,\AA, total
exposure time of 1.3 hours) gratings (see Figure \ref{fig:1851spec}). For NGC
6752, we used the HR4 (total exposure time of 3.6 hours) and HR19B (total
exposure time of 1.3 hours) gratings. A telluric standard was also observed
using the HR13 and HR19 gratings. 

For the purposes of this project, we decided that the higher spectral
resolution provided by the IFU mode relative to the MEDUSA mode (e.g., $R_{\rm
IFU}$ = 32,500 versus $R_{\rm MEDUSA}$ = 20,350 for HR4) was a major advantage
for deriving accurate and precise chemical abundances in the program stars.
Additionally, the smaller number of targets that can be observed in the IFU
mode (up to 15) versus MEDUSA mode (up to 132) was not considered to be a
disadvantage for this project. For NGC 6752, there were only some 21 objects
for which we had already derived N and O abundances and measurements of C
abundances in up to 15 of these targets would be sufficient to study the C+N+O
sum, i.e., an additional 100 measurements of C in objects with no N and O
measurements would be surplus to requirements. For NGC 1851, there were only
seven anomalous RGB objects brighter than $V$ $\simeq$ 14.8 such that the 15
targets could include a reasonable number of anomalous RGB, canonical RGB and
AGB stars. While our primary objective was to compare the canonical and
anomalous RGBs, recent studies of AGB stars have indicated that they may be
populated exclusively by Na-poor objects \citep{campbell13}. Furthermore, the
RGB and AGB populations in NGC 1851 exhibit a complex distribution in CN
molecular line strengths \citep{campbell12}. Therefore, we were also interested
in studying the CNO content of AGB stars in NGC 1851. 

For each program star, we examined the spectrum obtained from each spaxel.
After testing various options, we summed the four central spaxels (8, 9, 12 and
13) for a given star in a given observing block. In all cases, these four
spaxels contained the vast majority of the flux. To produce the final spectrum
for a program star, the spectra from multiple observing blocks were combined.
Note that star NGC 1851 NR 5246 is affected by TiO and we do not present
chemical abundances for this object. 

Heliocentric radial velocities for the NGC 1851 stars are presented in Table
\ref{tab:rv}. These values were determined by comparing the observed and rest
wavelengths for about 80 atomic lines in each program star. We find an average
value of +320.2 $\pm$ 1.3 \kms\ ($\sigma$ = 4.8 \kms) which agrees with values
from the literature: +320.5 $\pm$ 0.6 \kms\ \citep[][updated in
2010]{harris96} catalogue; 
+320.26 \kms\ (rms = 3.74 \kms) \citep{carretta11}; 
+320.0 $\pm$ 0.4 \kms\ \citep{scarpa11}; 
+318.2 $\pm$ 0.5 \kms\ \citep{gratton121851}; 
+319.5 $\pm$ 0.5 \kms\ \citep{marino14}.
That is, all program stars are likely cluster members. 

\begin{table}
 \centering
 \begin{minipage}{85mm}
  \caption{Heliocentric radial velocities for NGC 1851.} 
  \label{tab:rv} 
  \begin{tabular}{@{}lccc@{}}
  \hline
	Name & 
	CMD & 
	RV (\kms) & 
	$\sigma$RV (\kms) \\ 
        (1) & 
        (2) &
        (3) &
        (4) \\ 
  \hline
NR 712  & RGB & +315.4 & 0.5 \\
NR 1290 & RGB & +327.5 & 0.6 \\
NR 4740 & RGB & +317.6 & 0.5 \\
NR 6221 & RGB & +316.8 & 0.8 \\
NR 6250 & RGB & +322.6 & 0.7 \\
\\
NR 1469 & AGB & +313.3 & 0.8 \\
NR 2352 & AGB & +314.4 & 0.9 \\
NR 3272 & AGB & +326.4 & 0.7 \\
NR 8066 & AGB & +326.3 & 0.7 \\
\\
NR 2953 & m1  & +324.5 & 0.8 \\
NR 3213 & m1  & +321.8 & 0.7 \\
NR 5171 & m1  & +315.4 & 0.9 \\
NR 5246 & m1  & \ldots & \ldots \\
NR 5543 & m1  & +320.7 & 0.9 \\
NR 6217 & m1  & +321.6 & 0.8 \\
  \hline
\end{tabular}
\end{minipage}
\end{table}

\section{ANALYSIS}

We commenced our analysis using one-dimensional wavelength-calibrated pipeline
reduced spectra. The signal-to-noise ratio exceeded 100 per pixel for each
wavelength setting in all program stars. The spectra were normalised by fitting
low-order polynomial functions. For the HR19B spectra (@ $\sim$8000\,\AA), we
divided the program stars by the telluric standard. 

In the case of NGC 6752, stellar parameters for all program stars have been
obtained from our previous studies \citep{grundahl02,yong03}. In the case of
NGC 1851, we derived stellar parameters in the following manner (the approach
is very similar, but not identical, to that applied to NGC 6752). The effective
temperature, \teff, was determined using colour-temperature relations
\citep{alonso99,ramirez05} based on the infrared flux method. We used the
\strom\ photometry from \citet{grundahl99} and $JHK$ photometry from 2MASS
\citep{2mass} and adopted a reddening of $E(B-V)$ = 0.02 from the 2010 version
of the \citet{harris96} catalogue. For each star we obtained values for \teff\
from the two calibrations, \citet{alonso99} and \citet{ramirez05}, using the
$b-y$, $V-J$, $V-H$ and $V-K$ colours (with appropriate transformations from
2MASS to the $TCS$ system using the relations in \citealt{alonso94} and
\citealt{carpenter01}). We adopted the average \teff, weighted by the
uncertainties for each colour, and note that the mean difference,
\citeauthor{alonso99} $-$ \citeauthor{ramirez05}, is +55 K $\pm$ 17 K ($\sigma$
= 64 K). The surface gravity, \logg, was determined using \teff, a distance
modulus $(m-M)_V$ = 15.47 \citep{harris96}, bolometric corrections from
\citep{alonso99} and a mass of 0.8 \msun\ for the RGB objects and 0.7 \msun\
for the AGB stars. Observational constraints on the masses of RGB stars in
globular clusters can be obtained from eclipsing binaries near the main
sequence turnoff. In the comparable age and metallicity globular clusters 47
Tuc and M4, masses of $\sim$0.8 \msun\ are obtained by \citet{thompson10} and
\citet{kaluzny13}, respectively. AGB stars in globular clusters do not have
observational constraints on their masses, and these values are subject to
additional uncertainties including mass loss.  \citet{dotter08} estimate that
horizontal branch stars, i.e., the evolutionary phase prior to the AGB, have
lost $\sim$ 0.15 \msun, while \citet{gratton10} estimate a value of $\sim$ 0.20
\msun. Therefore, our assumption of 0.70 \msun\ on the AGB may be slightly
overestimated. 

We then measured equivalent widths (EWs) for a set of lines using {\sc
iraf}\footnote{{\sc iraf} is distributed by the National Optical Astronomy
Observatories, which are operated by the Association of Universities for
Research in Astronomy, Inc., under cooperative agreement with the National
Science Foundation.} and {\sc daospec} \citep{stetson08}. The line list is
presented in Table \ref{tab:ew}. With estimates for \teff\ and \logg\, we
generated one dimensional local thermodynamic equilibrium (LTE) model
atmospheres with [$\alpha$/Fe] = +0.4 from the \citet{castelli03} grid using
the interpolation software tested in \citet{allende04}.  Chemical abundances
were computed using the LTE stellar line analysis program {\sc moog}
\citep{moog,sobeck11}. The microturbulent velocity, \vt, was obtained by
forcing no trend between the abundance from \fei\ and the reduced equivalent
width, $\log ($EW$/\lambda)$. The average number of \fei\ and \feii\ lines
measured in a given star was 29 and 3, respectively. Following \citet{yong05},
we estimate that the internal uncertainties in \teff, \logg\ and \vt\ are 30 K,
0.1 dex and 0.1 \kms, respectively, for NGC 6752. For NGC 1851, the uncertainty
in \teff\ can be obtained by the weighted error from the colour temperature
relations. This value is 32 K, and we conservatively adopt 40 K as the
uncertainty. For the surface gravity, uncertainties in the temperature,
distance, reddening, \teff\ and $V$ mag when added in quadrature translate into
an error in \logg\ of 0.05 dex, and we conservatively adopt an error of 0.1
dex.  For \vt, we plotted this quantity versus \logg\ and fitted a straight
line to the data. The scatter about the linear fit was 0.17 \kms, and a similar
approach with respect to \teff\ resulted in a scatter of 0.18 \kms. Thus, we
adopt an uncertainty in \vt\ of 0.2 \kms. 

\begin{table}
 \centering
 \begin{minipage}{85mm}
  \caption{Line list for the NGC 1851 stars.} 
  \label{tab:ew} 
  \begin{tabular}{@{}cccccc@{}}
  \hline
	Wavelength & 
	Species\footnote{The digits to the left of the decimal point are the atomic
number. The digit to the right of the decimal point is the ionization state
(``0'' = neutral, ``1'' = singly ionised).} &
	L.E.P.\ & 
	$\log gf$ & 
	NR 712 & 
	Source\footnote{A = \citet{gratton03}; B = Oxford group including 
\citet{blackwell79feb,blackwell79fea,blackwell80fea,blackwell86fea,blackwell95fea}.
\\
This table is published in its entirety in the electronic edition of the paper.
A portion is shown here for guidance regarding its form and content.}\\
        \AA  & 
             &
        (eV) &
             & 
        (m\AA)  & 
            \\ 
        (1) & 
        (2) &
        (3) &
        (4) & 
        (5) & 
        (6) \\ 
  \hline
6154.23 & 11.0 & 2.10 & $-$1.57 &   11.6 &  A \\
6160.75 & 11.0 & 2.10 & $-$1.26 &   18.7 &  A \\
6318.71 & 12.0 & 5.11 & $-$1.94 &   25.3 &  A \\
6319.24 & 12.0 & 5.11 & $-$2.16 & \ldots &  A \\
6120.24 & 26.0 & 0.91 & $-$5.97 &   19.0 &  B \\
6136.62 & 26.0 & 2.45 & $-$1.40 & \ldots &  B \\
6151.62 & 26.0 & 2.18 & $-$3.30 &   66.7 &  A \\
  \hline
\end{tabular}
\end{minipage}
\end{table}

Carbon abundances were obtained by comparing synthetic with observed spectra in
the vicinity of the 4300\,\AA\ CH molecular lines. We used the CH line list
compiled by B.\ Plez et al.\ (2009, private communication). In our analysis,
the dissociation energy for CH was 3.465 eV. 

Although nitrogen abundances had already been obtained in NGC 6752 based on the
3360\,\AA\ NH molecular lines \citep{yong08nh}, we re-measured these values using
the \citet{sobeck11} version of {\sc moog}. The average difference (2008 values
minus updated values) is +0.09 dex $\pm$ 0.01 dex ($\sigma$ = 0.05 dex). We
also measured nitrogen from the 8000\,\AA\ CN molecular lines using the line list
from \citet{reddy02}. The dissociation energy for CN was 7.750 eV. (When using a
CN line list kindly provided by M.\ Asplund, essentially identical N abundances
were obtained.) 

Oxygen abundances were determined by comparing synthetic and observed spectra
near the 6300\,\AA\ [OI] line. In NGC 1851, relative abundances for Na, Mg and
Zr were measured based on an equivalent-width analysis. The \citet{asplund09}
solar abundances were adopted and the chemical abundances are presented in
Tables \ref{tab:6752}, \ref{tab:1851} and \ref{tab:1851namgzr}. The
metallicity, [Fe/H], was determined by averaging the results from \fei\ and
\feii\ weighted by the number of lines from each species (this approach
strongly favours \fei). 

As noted in the introduction, the abundances for C, N and O are coupled due
to molecular equilibrium. The processes described above required iteration
until self-consistent abundances were obtained for a given program star. In the
upper panel of Figure \ref{fig:1851spec}, the star with the lower C abundance
(NR 2953) has stronger CH lines relative to the star with the higher C abundance
(NR 6250). This apparently unusual situation is a direct consequence of the
relative O abundances (NR 6250 has a considerably higher abundance) and
molecular equilibrium. 

Uncertainties in the abundance ratios were obtained in the following manner. We
repeated the analysis and varied the stellar parameters, one at a time, by
their uncertainties. We also considered the uncertainty in the metallicity used
to generate the model atmosphere, [m/H], and varied this value by 0.1 dex. The
systematic uncertainty was obtained by adding these four error terms, in
quadrature (although we note that this approach ignores covariances). To obtain
the total error, we added the systematic and random errors in quadrature. Due
to molecular equilibrium, the uncertainty in the O abundance affects the
derived C abundance, and vice versa. For these two species, we include an
additional error term accounting for the uncertainty in these abundances. For
the N abundances derived from the CN molecular lines, we also need to take into
account the uncertainties in the C and O abundances.  The uncertainty in the O
abundance produces the dominant term in the error budget for N as derived from
CN lines. For these CNO abundances derived from fitting synthetic spectra, we
adopted a fitting error based on $\chi^2$ analysis (i.e., the value for which
$\Delta\chi^2$ = 1) and use these values as the random error. The errors are
presented in Table \ref{tab:parvar}. 

\begin{table*}
 \centering
 \begin{minipage}{120mm}
  \caption{Na, Mg and Zr abundances for NGC 1851.} 
  \label{tab:1851namgzr} 
  \begin{tabular}{@{}lcccrrcr@{}}
  \hline
	Name & 
	CMD & 
	A(Na) &
	A(Mg) &
	A(Zr) & 
	[Na$_{\rm NLTE}$/Fe]\footnote{Non-LTE corrections from \citet{lind11}.} &
	[Mg/Fe] & 
	[Zr/Fe] \\ 
        (1) & 
        (2) &
        (3) &
	(4) & 
        (5) & 
        (6) & 
        (7) & 
        (8) \\ 
  \hline
NR 712  & RGB &    4.79 &    6.57 &    1.35 & $-$0.18 &    0.30 &    0.10 \\ 
NR 1290 & RGB &    4.99 &    6.63 &    1.38 & $-$0.04 &    0.29 &    0.07 \\ 
NR 4740 & RGB &    4.78 &    6.60 &    1.33 & $-$0.31 &    0.20 & $-$0.05 \\ 
NR 6221 & RGB &    5.01 &    6.59 &    1.49 & $-$0.02 &    0.26 &    0.18 \\ 
NR 6250 & RGB &    5.06 &    6.55 &    1.37 &    0.07 &    0.25 &    0.09 \\ 
\\
NR 1469 & AGB &    5.35 &    6.56 & $<$1.73 &    0.35 &    0.28 & $<$0.47 \\ 
NR 2352 & AGB &    5.34 &    6.70 & $<$1.83 &    0.18 &    0.26 & $<$0.41 \\ 
NR 3272 & AGB &    5.04 &    6.43 & $<$1.67 &    0.06 &    0.16 & $<$0.42 \\ 
NR 8066 & AGB &    5.38 &    6.51 & $<$1.66 &    0.39 &    0.22 & $<$0.40 \\ 
\\
NR 2953 & m1  &    5.74 &    6.75 &    1.61 &    0.60 &    0.36 &    0.23 \\ 
NR 3213 & m1  &    5.73 &    6.74 &    1.81 &    0.70 &    0.44 &    0.53 \\ 
NR 5171 & m1  &    5.34 &    6.61 &    1.51 &    0.37 &    0.35 &    0.27 \\ 
NR 5246 & m1  &  \ldots &  \ldots &  \ldots &  \ldots &  \ldots &  \ldots \\ 
NR 5543 & m1  &    5.47 &    6.70 &    1.59 &    0.46 &    0.40 &    0.31 \\ 
NR 6217 & m1  &    5.46 &    6.73 &    1.45 &    0.41 &    0.39 &    0.13 \\ 
  \hline
\end{tabular}
\end{minipage}
\end{table*}

\begin{table*}
 \centering
 \begin{minipage}{180mm}
  \caption{Abundance errors from uncertainties in atmospheric parameters and
element abundances.} 
  \label{tab:parvar} 
  \begin{tabular}{@{}lrrrrrrrr@{}}
  \hline
	Name & 
	C & 
	N$_{\rm NH}$ &
	N$_{\rm CN}$ &
	O &
	Na &
	Mg &
	Zr & 
        Fe \\ 
        (1) & 
        (2) &
        (3) &
	(4) & 
	(5) & 
	(6) & 
	(7) & 
        (8) & 
        (9) \\ 
  \hline
\multicolumn{9}{c}{NGC 6752-11} \\ 
\teff\ + 30 K      & $-$0.03 & $-$0.04 & $-$0.05 & $-$0.01 &  \ldots &  \ldots &  \ldots &  \ldots \\ 
\logg\ + 0.1 dex   &    0.01 &    0.04 &    0.00 & $-$0.04 &  \ldots &  \ldots &  \ldots &  \ldots \\ 
\vt\ + 0.1 \kms    &    0.02 &    0.04 &    0.01 &    0.00 &  \ldots &  \ldots &  \ldots &  \ldots \\ 
$\textrm{[m/H]}$ + 0.1 dex & $-$0.04 & $-$0.03 & $-$0.05 & $-$0.03 &  \ldots &  \ldots &  \ldots &  \ldots \\ 
A(O) + 0.05 dex    & $-$0.01 &  \ldots &    0.07 &  \ldots &  \ldots &  \ldots &  \ldots &  \ldots \\ 
A(C) + 0.06 dex    &  \ldots &  \ldots & $-$0.01 &    0.00 &  \ldots &  \ldots &  \ldots &  \ldots \\ 
Random error\footnote{For C, N and O, this is the fitting error based on
$\chi^2$ analysis. For other elements, this is the standard error of the mean.} &    0.04 &    0.05 &    0.03 &    0.03 &  \ldots &  \ldots &  \ldots &  \ldots \\ 
Total              &    0.07 &    0.09 &    0.10 &    0.06 &  \ldots &  \ldots &  \ldots &  \ldots \\ 
\hline
\multicolumn{9}{c}{NGC 1851 NR 4740 (RGB)} \\ 
\teff\ + 40 K      & $-$0.04 &  \ldots &    0.03 &    0.01 & $-$0.04 & $-$0.01 & $-$0.09 & $-$0.02 \\ 
\logg\ + 0.1 dex   & $-$0.04 &  \ldots & $-$0.04 & $-$0.04 &    0.01 &    0.00 &    0.00 & $-$0.03 \\ 
\vt\ + 0.2 \kms    & $-$0.02 &  \ldots &    0.01 &    0.02 &    0.02 &    0.02 &    0.01 &    0.06 \\ 
$\textrm{[m/H]}$ + 0.1 dex & $-$0.07 &  \ldots & $-$0.03 & $-$0.02 &    0.00 &    0.00 &    0.00 & $-$0.01 \\ 
A(O) + 0.05 dex    & $-$0.04 &  \ldots &    0.10 &  \ldots &  \ldots &  \ldots &  \ldots &  \ldots \\ 
A(C) + 0.09 dex    &  \ldots &  \ldots & $-$0.04 & $-$0.02 &  \ldots &  \ldots &  \ldots &  \ldots \\ 
Random error       &    0.04 &  \ldots &    0.03 &    0.03 &    0.04 &    0.05 &    0.04 &    0.04 \\ 
Total              &    0.11 &  \ldots &    0.13 &    0.06 &    0.06 &    0.06 &    0.10 &    0.08 \\ 
\hline 
\multicolumn{9}{c}{NGC 1851 NR 6217 (m1)} \\ 
\teff\ + 40 K      & $-$0.01 &  \ldots &    0.05 &    0.01 & $-$0.03 &    0.01 & $-$0.08 &    0.01 \\ 
\logg\ + 0.1 dex   & $-$0.01 &  \ldots & $-$0.08 & $-$0.05 &    0.00 &    0.00 &    0.00 & $-$0.03 \\ 
\vt\ + 0.2 \kms    &    0.02 &  \ldots &    0.02 &    0.01 &    0.04 &    0.02 &    0.02 &    0.05 \\ 
$\textrm{[m/H]}$ + 0.1 dex & $-$0.11 &  \ldots & $-$0.06 & $-$0.01 &    0.01 &    0.00 &    0.00 & $-$0.01 \\ 
A(O) + 0.05 dex    & $-$0.02 &  \ldots &    0.18 &  \ldots &  \ldots &  \ldots &  \ldots &  \ldots \\ 
A(C) + 0.11 dex    &  \ldots &  \ldots & $-$0.05 &    0.01 &  \ldots &  \ldots &  \ldots &  \ldots \\ 
Random error       &    0.04 &  \ldots &    0.03 &    0.04 &    0.06 &    0.05 &    0.06 &    0.05 \\ 
Total              &    0.12 &  \ldots &    0.22 &    0.07 &    0.08 &    0.05 &    0.10 &    0.08 \\ 
  \hline
\end{tabular}
\end{minipage}
\end{table*}

For NGC 1851, a subset of our program stars were studied in
\citet{carretta101851}, \citet{carretta11} and \citet{villanova10}. We defer
our comparison with the latter until Section 4.2. For the seven stars in common
with \citet{carretta101851} and \citet{carretta11}, we find the following
differences in the sense ``this study $-$ \citeauthor{carretta101851}'': 
$\Delta$RV = +0.8 $\pm$ 0.3 \kms;
$\Delta$\teff = $-$10 $\pm$ 14 K;
$\Delta$\logg = $-$0.05 $\pm$ 0.01 cgs; 
$\Delta$\vt = +0.20 $\pm$ 0.14 \kms;
$\Delta$[Fe/H] = $-$0.14 $\pm$ 0.03 dex;
$\Delta$[O/Fe] = +0.11 $\pm$ 0.16 dex;
$\Delta$[Na/Fe] = $-$0.15 $\pm$ 0.06 dex;
$\Delta$[Mg/Fe] = $-$0.04 $\pm$ 0.05 dex;
$\Delta$[Zr/Fe] = $-$0.03 $\pm$ 0.09 dex. 
Overall, our radial velocities, stellar parameters and chemical abundances are
in good agreement.

\section{RESULTS AND DISCUSSION}

\subsection{NGC 6752}

\begin{figure}
\centering
      \includegraphics[width=.75\hsize]{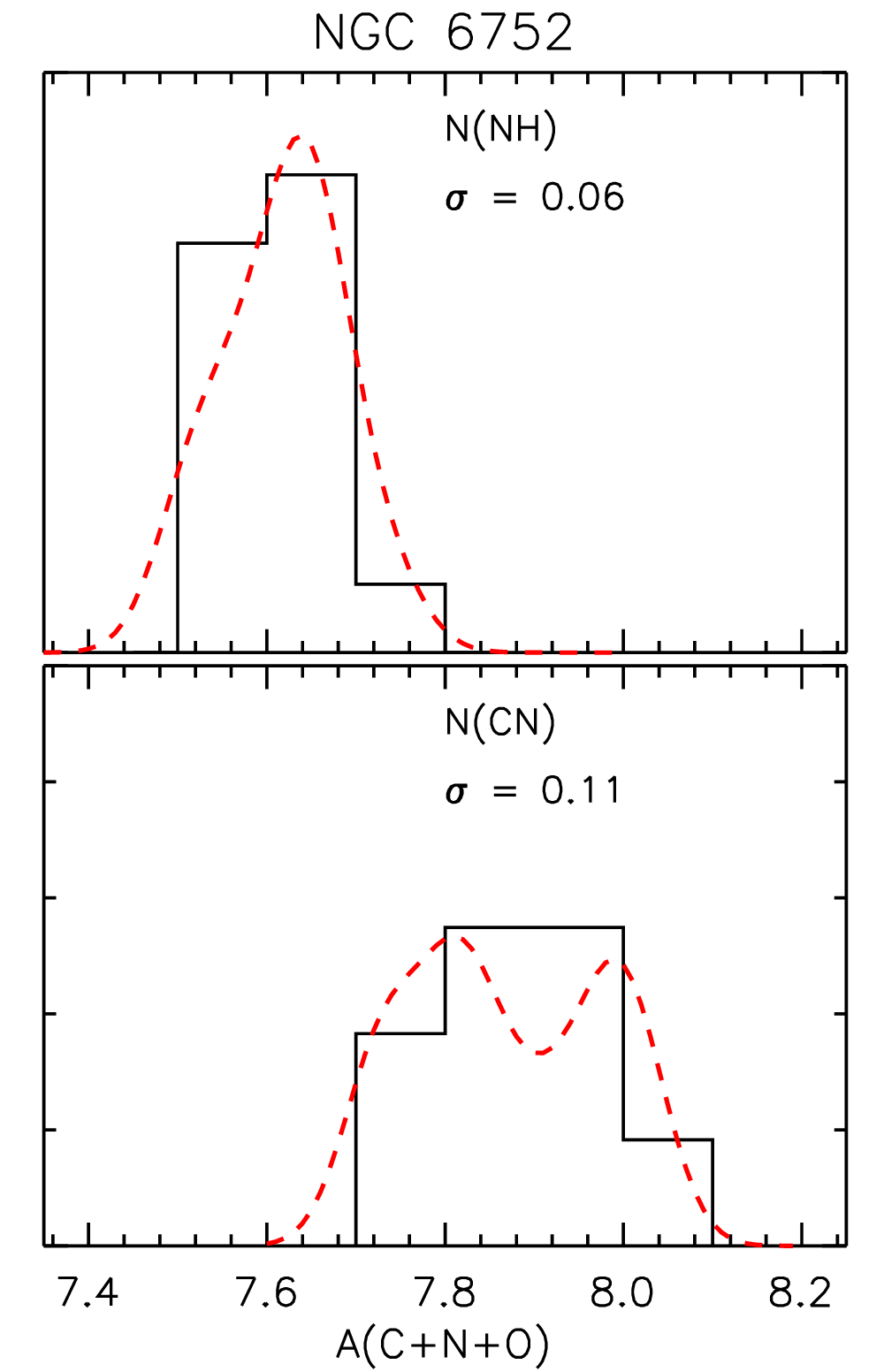}
      \caption{The distribution of the C+N+O abundance sum in NGC 6752 (the
black histogram has a bin width of 0.1 dex). The red dashed line is a
generalised histogram (Gaussian kernel $\sigma$ = 0.10 dex). In the upper
panel, the N abundances are from analysis of the NH lines. In the lower panel,
the N abundances are from analysis of the CN lines. 
      \label{fig:6752cno} }
\end{figure}

In Figure \ref{fig:6752cno}, we plot the C+N+O distribution for NGC 6752. In
the upper panel, the N abundances are derived from analysis of the NH lines.
The mean C+N+O abundance is 7.62 $\pm$ 0.02 dex and the standard deviation of
the C+N+O distribution is $\sigma$ = 0.06 $\pm$ 0.01 dex. In order to
understand whether this abundance dispersion is consistent with a constant
C+N+O value convolved with the measurement uncertainty, we conducted the
following test. For a representative star, we replaced the C abundance by a
random number drawn from a normal distribution of width 0.07 dex (i.e., the
measurement uncertainty for C), centered at the $\log\epsilon$ (C) value. A
similar approach was taken to replace the N and O abundances, and we note that
their measurement uncertainties are 0.09 and 0.06 dex, respectively. This
produces a new set of C+N+O abundances. We repeated the process for 10$^{6}$
realisations and measured the standard deviation of the C+N+O distribution.  As
expected, when considering a star whose C+N+O sum is dominated by N, the
standard deviation of the C+N+O distribution is 0.08 dex, and this value is
essentially the uncertainty in N, 0.09 dex. Similarly, for a star whose C+N+O
sum is dominated by O, the standard deviation of the C+N+O distribution is 0.05
dex, and this is comparable to the uncertainty in O, 0.06 dex. Therefore, we
argue that our C+N+O distribution is consistent with a single value convolved
with the measurement errors. 
  
In the lower panel of Figure \ref{fig:6752cno}, we again plot the C+N+O
distribution but with N abundances derived from analysis of the CN lines. In
this case, the mean C+N+O abundance is 7.87 $\pm$ 0.04 dex and the standard
deviation of the C+N+O distribution is $\sigma$ = 0.11 $\pm$ 0.03 dex. To
understand whether the abundance dispersion is consistent with a constant value
of C+N+O convolved with the measurement uncertainty, we adopted the same
approach described above but with the uncertainty in N of 0.10 dex as
appropriate for the CN analysis. As before, we find that the C+N+O
distribution is consistent with a single value when taking into account the
measurement errors. Therefore, the first main conclusion we draw is that the
C+N+O abundance sum in NGC 6752 is constant. While a similar conclusion was
reached by \citet{carretta05}, in this work we achieve higher precision; our
errors in the C+N+O sum are at or below the 0.10 dex level. 

\begin{figure}
\centering
      \includegraphics[width=.99\hsize]{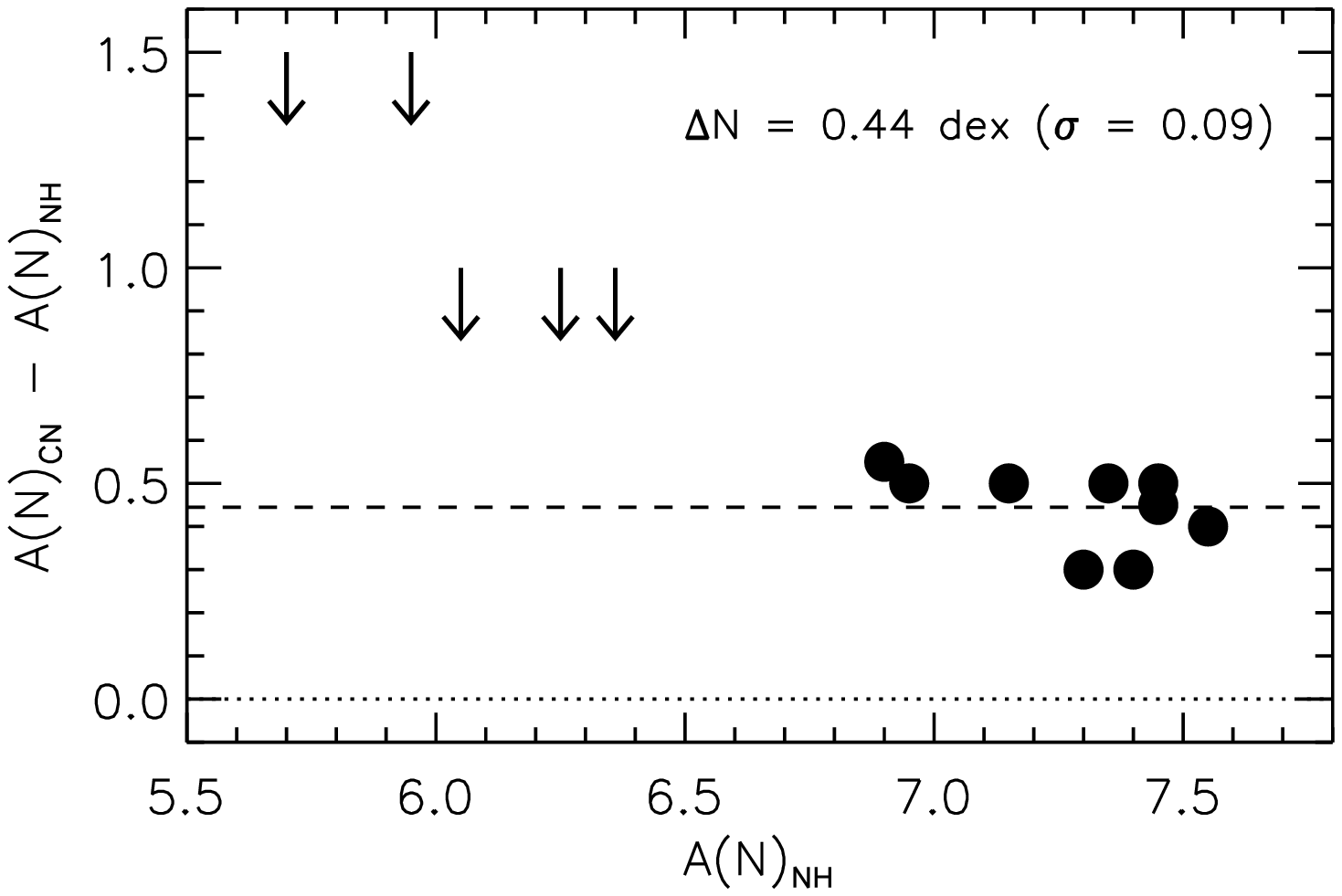}
      \caption{The difference in N abundance from the CN and NH molecular lines
versus N abundance (NH) in NGC 6752. 
      \label{fig:ndiff} }
\end{figure}

It is also evident that there is a systematic difference in the N abundance
derived from the different molecular lines, NH versus CN. In Figure
\ref{fig:ndiff}, we plot the N abundance difference and note that the N
abundance as derived from the CN molecular lines exceeds the values from the NH
molecular lines by an average of 0.44 dex ($\sigma$ = 0.09 dex). While the
difference in N abundance directly affects the C+N+O abundance sum, the reason
for this zero-point offset is not obvious. In their study of metal-poor giant
stars, \citet{spite05} measured N abundances using the 3360 \AA\ NH lines and
the 3890 \AA\ CN lines and found a 0.4 dex offset. In their case, the N
abundances from NH exceeded those from CN. They attributed the abundance
differences to uncertainties in the line positions, $gf$ values and
dissociation energy, and it is likely that a similar explanation applies to the
N abundance offset in this study. We adopt the N abundance as derived from NH
since this quantity has no dependence on the C and O abundances. 

In light of the systematic difference in N abundance, we may ask the following
question: if we expect that the C+N+O abundance sum should be constant, what
would be the systematic shift in N abundances (as derived from NH) that
produces the smallest abundance dispersion for C+N+O? The answer is a shift of
+0.08 dex. When such an arbitrary shift is made, the resulting C+N+O abundance
dispersion is essentially identical to our 0.06 dex value. The systematic N
abundance differences underscore the importance of zero-point offsets when
determining abundance sums such as C+N+O. 

The program stars were selected by \citet{grundahl02} to lie above and below
the RGB bump, $V$ = 13.626 \citep{nataf13}. When adopting the N abundances from
NH, the average C+N+O values for stars brighter and fainter than the RGB bump
are identical, 7.62 $\pm$ 0.03 ($\sigma$ = 0.07). 

\subsection{NGC 1851} 

We commence by noting that the N abundances in NGC 1851 were derived from
analysis of the CN lines. Recall that for NGC 6752, there was a 0.44 dex
systematic offset between the N abundances from NH and CN. We adopted the NH
values for NGC 6752. Therefore, when computing the C+N+O abundance sum for NGC
1851, we adjust the N abundances by $-$0.44 dex to place the two clusters on
the same scale\footnote{We are assuming that the offset inferred from NGC 6752
is applicable to NGC 1851. Clearly it would be of interest to measure N from NH
in NGC 1851.}. 

\begin{figure}
\centering
      \includegraphics[width=.75\hsize]{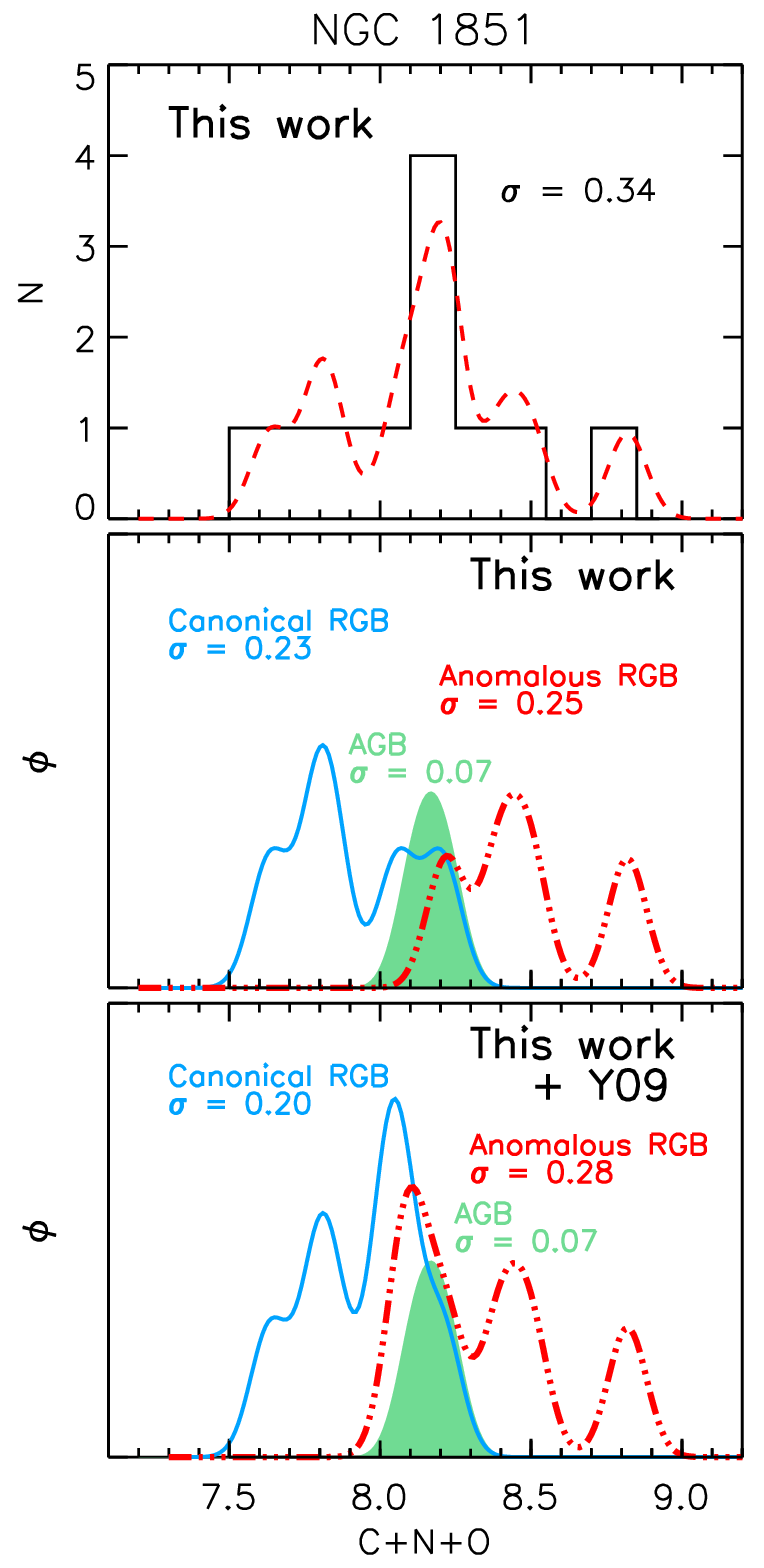}
      \caption{The distribution of the C+N+O abundance sum in NGC 1851 (the
black histogram has a bin width of 0.15 dex). The red dashed line is a
generalised histogram (Gaussian kernel $\sigma$ = 0.15 dex). In the middle
panel, generalised histograms are presented for the canonical RGB (5 stars),
anomalous RGB (4 stars) and the AGB (2 stars). The lower panel includes data
from \citet{yong09}. 
      \label{fig:1851cno} }
\end{figure}

In Figure \ref{fig:1851cno}, we plot the C+N+O abundance distribution for NGC
1851. The mean C+N+O abundance is 8.16 $\pm$ 0.10 dex and the C+N+O abundance
distribution is broad; the standard deviation is $\sigma$ = 0.34 $\pm$ 0.08 dex
and the values span more than a factor of 10. To understand whether the
observed abundance distribution is consistent with no intrinsic abundance
dispersion, we adopted the same approach as for NGC 6752. For a representative
canonical RGB star, our C, N and O uncertainties are 0.11, 0.13 and 0.06 dex,
respectively. For a given canonical RGB star, we updated each of the C, N and O
abundances by drawing random numbers from normal distributions of widths
corresponding to the appropriate uncertainties and generated new C+N+O
abundances. We repeated the process for 10$^{6}$ realisations and measured the
standard deviation of the C+N+O distribution. For the five canonical RGB
objects, the standard deviations ranged from 0.06 to 0.10 dex, and as in the
case of NGC 6752, these values depend on whether the CNO content is dominated
by N or O. We repeated the process for the anomalous RGB objects noting that
for a representative star, the C, N and O uncertainties are 0.12, 0.22 and 0.06
dex, respectively. For the four anomalous RGB objects, the standard deviations
of the C+N+O distribution (based on 10$^{6}$ realisations) ranged from 0.19 to
0.21 dex (and we note that for all stars N dominates the C+N+O sum). In light
of this error analysis, the C+N+O abundance distribution ($\sigma$ = 0.34 $\pm$
0.08 dex) plotted in the upper panel of Figure \ref{fig:1851cno} appears to be
inconsistent with a single C+N+O value convolved with measurement uncertainties
($\lesssim$ 0.20 dex). 
 
In the middle panel of Figure \ref{fig:1851cno}, we plot the C+N+O
distributions for the canonical RGB, AGB and anomalous RGB populations. The
mean C+N+O abundances for each of the canonical and anomalous RGB populations
are 7.90 $\pm$ 0.10 dex ($\sigma$ = 0.23 $\pm$ 0.08 dex) and 8.48 $\pm$ 0.13
dex ($\sigma$ = 0.25 $\pm$ 0.09 dex), respectively. We therefore confirm
differences in CNO content between the canonical and anomalous RGB samples in
NGC 1851, and this is the second main result in this paper. 

Setting aside the two AGB stars, the canonical and anomalous RGB populations
exhibit standard deviations for C+N+O of $\sigma$ = 0.23 $\pm$ 0.08 and
$\sigma$ = 0.25 $\pm$ 0.09~dex, respectively. In the case of the canonical RGB
sample, the distribution is broader than that expected from measurement
uncertainties alone ($\lesssim$ 0.10~dex) and therefore indicates that there
may be an intrinsic C+N+O spread within the canonical RGB population. For
comparison, recall that in NGC 6752 (for both the CN and NH analyses), the
C+N+O distribution exhibited no evidence for an intrinsic abundance spread
given the measurement uncertainty ($\le$ 0.11 dex). In the case of the
anomalous RGB sample in NGC 1851, the C+N+O distribution may be consistent with
a constant value combined with the measurement uncertainties ($\sim$~0.20 dex).
We emphasise, however, that our sample sizes for both the canonical and
anomalous RGBs are small, and therefore larger samples are needed to explore
whether or not each population hosts an intrinsic spread in C+N+O.  We also
note that \citet{carretta141851} found that the anomalous RGB is populated
almost exclusively by N-rich stars. Within our limited sample, the anomalous
RGB objects are all more N rich with respect to the canonical RGB. 

We previously published C+N+O values for four stars in NGC 1851 \citep{yong09}
using the same spectral features as in this study. In the lower panel of Figure
\ref{fig:1851cno}, we combine those values with the current work (shifting the
N abundances by $-$0.44 dex to be consistent with this study). Inclusion of
those four stars (two in each of the canonical and anomalous RGBs) does not
change the two key results, namely, that the mean C+N+O abundance is higher for
the anomalous RGB (8.35 $\pm$ 0.11 dex) compared to the canonical RGB (7.94
$\pm$ 0.07 dex) and that the dispersion in C+N+O for the canonical RGB
($\sigma$ = 0.28 $\pm$ 0.09 dex) likely exceeds that expected from the
measurement uncertainties. The dispersion for the anomalous RGB, $\sigma$ =
0.20 $\pm$ 0.06 dex, can be attributed to the measurement uncertainties. 

Had we not applied the 0.44 dex shift to the N abundances, the canonical RGB
would still have lower C+N+O compared to the anomalous RGB. Regardless of
whether we include the \citet{yong09} sample or apply an abundance correction
to N, in all cases the anomalous RGB has a higher content of C+N+O compared to
the canonical RGB. Such a result supports the scenario proposed by
\citet{cassisi08} in which the two subgiant branch populations are roughly
coeval, but with different C+N+O abundances. 

On the other hand, \citet{villanova10} reported constant C+N+O abundances for a
sample of 15 red giants in NGC 1851. Their sample consisted of eight and seven
stars on the canonical and anomalous RGBs, respectively. They used the same
diagnostics to measure the CNO abundances as in this study and obtained C+N+O
values of 7.99 $\pm$ 0.02 ($\sigma$ = 0.07) and 8.02 $\pm$ 0.04 ($\sigma$ =
0.11) for the canonical and anomalous RGBs, respectively. For comparison, our
values for the canonical and anomalous RGBs are 7.90 $\pm$ 0.10 and 8.48 $\pm$
0.13 dex, respectively. For the canonical RGB, our C+N+O values are in
agreement. For the anomalous RGB, our C+N+O values disagree by $\sim$ 0.45 dex. 

There are three stars in common between this study and \citet{villanova10}: NR
3213 = ID 9; NR 5543 = ID 16; NR 6217 = ID 20. For quantities published by both
studies, we examine the differences in the sense ``this study $-$
\citeauthor{villanova10}'' (while [Fe/H] can be compared for all three stars,
we only measured O and C+N+O for the latter two objects) and find the
following: $\Delta$[Fe/H] = $-$0.05 $\pm$ 0.02; $\Delta$A(O) = $-$0.04 $\pm$
0.08; $\Delta$C+N+O = +0.42 $\pm$ 0.10 dex. We are unable to compare stellar
parameters (\teff, \logg, \vt), radial velocities or individual C and N
abundances since \citet{villanova10} did not publish these values.
Nevertheless, there is good agreement for [Fe/H] and A(O). The stars in common
are N-rich, so the C+N+O differences between this work and their study are
likely due to differences in N abundance. 

It is important to recognise, however, that we are analysing RGB objects rather
than subgiant branch stars. Therefore any conclusions we draw concerning the
CNO content of subgiant branch stars in NGC 1851 will necessarily assume that
the abundances we derive for RGB objects would be similar to those on the
subgiant branch. That said, we can compare our average abundances for the two
RGBs to measurements of subgiant branch stars by \citet{lardo121851}. They
measured C and N (but not O) in subgiant branch stars in NGC 1851 and found
that the fainter subgiant branch had a higher C+N content than the brighter
subgiant branch, 7.64 $\pm$ 0.24 and 7.23 $\pm$ 0.31, respectively. Given that
the fainter subgiant branch connects to the anomalous RGB, our C+N values for
the anomalous (8.45 $\pm$ 0.14) and canonical (7.52 $\pm$ 0.21) RGBs are
qualitatively consistent with \citet{lardo121851}, although we note that they
used different diagnostics to measure N abundances compared to this study. 

\begin{figure}
\centering
      \includegraphics[width=.99\hsize]{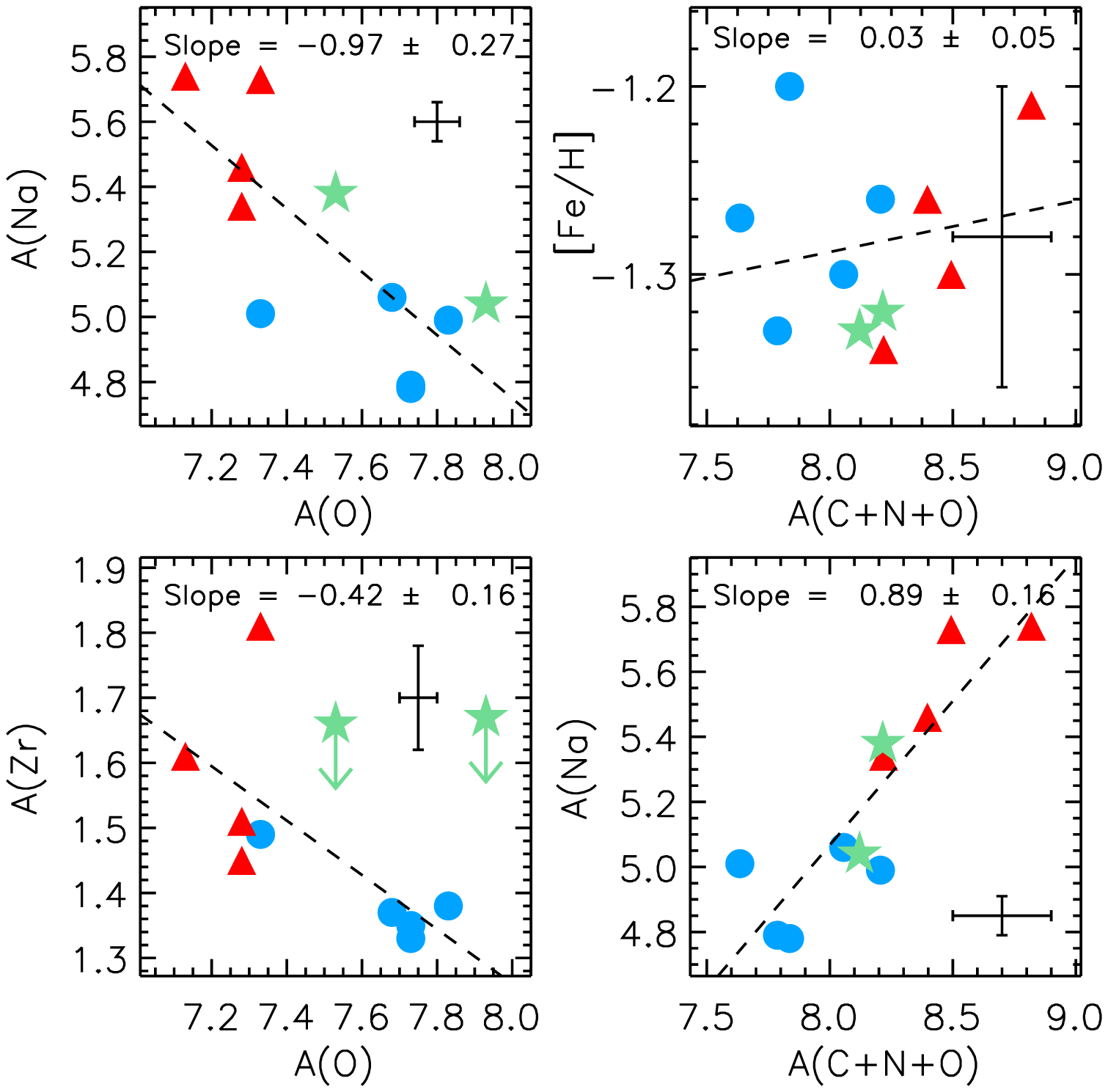} 
      \caption{Abundance ratios for combinations of the light elements (C, N,
O, Na), Fe and Zr in NGC 1851. The dashed line is the linear fit to the data,
excluding limits, (slope and error are included in each panel).  A
representative error bar is included in each panel. 
      \label{fig:1851namgzr} }
\end{figure}

We now briefly discuss the two AGB stars. With or without the arbitrary shift
in N abundance, these two objects have C+N+O values that lie between the
canonical and anomalous RGB populations. As no change in the C+N+O sum is
expected to take place between the RGB and AGB (e.g., see \citealt{karakas14}
and references therein), the AGBs could come from either the upper envelope of
the canonical RGB C+N+O distribution or from the lower envelope of the
anomalous RGB C+N+O distribution. Given the known differences in
neutron-capture element abundances between the canonical and anomalous RGBs in
NGC 1851 \citep{yong081851,villanova10,carretta11}, measurements of
neutron-capture element abundances in the AGB stars could reveal whether they
are chemically related to a particular RGB. Our Zr measurements in the
canonical and anomalous RGBs follow the established pattern in this cluster,
i.e., the average Zr abundance in the anomalous RGB population is 0.21 $\pm$
0.08 dex higher than in the canonical RGB sample (see Figure
\ref{fig:1851namgzr}). Unfortunately, we could only obtain upper limits to the
Zr abundance for all AGB stars, and those limits could be consistent with
either the Zr-rich anomalous RGB or the Zr-normal canonical RGB. Given the
modest wavelength coverage for the NGC 1851 sample, we were unable to identify
lines of neutron-capture elements that would yield reliable abundance
measurements. Figure \ref{fig:1851namgzr} indicates that the canonical and
anomalous RGBs have distinct O abundances. The two AGB stars have O abundances
in accord with the anomalous RGB, although any suggestion of association would
be speculation given the small numbers of stars. 

As noted in the introduction, there is evidence for a small iron abundance
dispersion in NGC 1851 \citep{carretta101851,carretta11}. For our [Fe/H]
measurements, the standard deviation is 0.055 $\pm$ 0.011 dex. While
\citet{carretta11} obtained a similar value, 0.051 $\pm$ 0.005 dex, our
measurement errors are 0.08 dex (see Table \ref{tab:parvar}) such that the
dispersion can be explained entirely by the measurement uncertainties. In
Figure \ref{fig:1851namgzr}, there is no correlation between metallicity,
[Fe/H], and C+N+O. In this figure, we also confirm the anticorrelation between
O and Na \citep{carretta101851,carretta11}. Additionally, we identify a
positive correlation between Na and C+N+O. 

Finally, \citet{marino11} examined CNO abundances in the globular cluster M22.
Like NGC 1851, M22 possesses a double subgiant branch as well as a spread in
$s$-process element abundances. \citet{marino11} found that the $s$-process
rich stars (which preferentially populate the fainter subgiant branch and
anomalous RGB) have a higher C+N+O content compared to the $s$-process normal
stars, 7.84 $\pm$ 0.03 ($\sigma$ = 0.07) and 7.57 $\pm$ 0.03 ($\sigma$ = 0.09),
respectively. We stress, however, that M22 and NGC 1851 are rather different
objects with distinct mean metallicities, metallicity dispersions, absolute
luminosities and kinematics \citep{dinescu97,casetti-dinescu13}. 

\section{CONCLUDING REMARKS}

We have studied the C+N+O abundance sum in the globular clusters NGC 6752 and
NGC 1851. For NGC 6752, there is no evidence for an intrinsic abundance
dispersion given the measurement uncertainties ($\lesssim$ 0.10 dex), although
the absolute value of the C+N+O sum depends on which set of molecular lines (NH
versus CN) are used to obtain the N abundance. While such a result confirms
previous investigations of this cluster, this study imposes considerably
tighter constraints on the source of the light element abundance variations.
The AGBs, fast-rotating massive stars and/or massive binaries that may have
operated in the early life of this cluster to produce the abundance variations
for O, Na etc.\ must not alter the C+N+O sum. If NGC 6752 is representative of
the least complex globular clusters, then by extension all globular clusters
that exhibit no evidence for a metallicity variation or multiple subgiant
branches may also have a constant C+N+O abundance sum despite large variations
for individual light element abundances. 

For NGC 1851, we confirm a large dispersion in the C+N+O abundance sum. That is
to say, the observed C+N+O dispersion ($\sigma$ = 0.34 $\pm$ 0.08 dex) far
exceeds that expected from measurement uncertainties alone ($\sim$ 0.20 dex).
We find that the anomalous RGB has a higher C+N+O content than the canonical
RGB by a factor of $\sim$ 0.6 dex. Such a result would support the
scenario in which the two subgiant branch populations are roughly coeval, but
with a different C+N+O abundance sum. Within the limited sample of canonical
RGB objects, there is evidence that the C+N+O abundance dispersion exceeds the
measurement uncertainties and this may indicate an intrinsic spread within this
population. Confirming such an abundance dispersion within the canonical RGB
population in this cluster would be of great interest for understanding the
formation history of this complicated object. 

\section*{Acknowledgments}

We thank Martin Asplund, Thibaut Decressin, Aaron Dotter, Amanda Karakas and
Anna F.\ Marino for helpful discussions. 
We thank Karin Lind and Sandro Villanova for providing data. 
We thank the anonymous referee for helpful comments. 
D.Y and J.E.N gratefully acknowledge support from the Australian Research
Council (grants DP0984924 and DP120101237). 
Funding for the Stellar Astrophysics Centre is provided by The Danish National
Research Foundation. The research is supported by the ASTERISK project
(ASTERoseismic Investigations with SONG and Kepler) funded by the European
Research Council (Grant agreement no.: 267864).

\label{lastpage}


\begin{thebibliography}{108}
\expandafter\ifx\csname natexlab\endcsname\relax\def\natexlab#1{#1}\fi

\bibitem[{{Allende Prieto} {et~al.}(2004){Allende Prieto}, {Barklem},
  {Lambert}, \& {Cunha}}]{allende04}
{Allende Prieto}, C., {Barklem}, P.~S., {Lambert}, D.~L., \& {Cunha}, K. 2004,
  \aap, 420, 183

\bibitem[{{Alonso} {et~al.}(1994){Alonso}, {Arribas}, \&
  {Martinez-Roger}}]{alonso94}
{Alonso}, A., {Arribas}, S., \& {Martinez-Roger}, C. 1994, \aaps, 107, 365

\bibitem[{{Alonso} {et~al.}(1999){Alonso}, {Arribas}, \&
  {Mart{\'{\i}}nez-Roger}}]{alonso99}
{Alonso}, A., {Arribas}, S., \& {Mart{\'{\i}}nez-Roger}, C. 1999, \aaps, 140,
  261

\bibitem[{{Asplund} {et~al.}(2009){Asplund}, {Grevesse}, {Sauval}, \&
  {Scott}}]{asplund09}
{Asplund}, M., {Grevesse}, N., {Sauval}, A.~J., \& {Scott}, P. 2009, \araa, 47,
  481

\bibitem[{{Bastian} {et~al.}(2013){Bastian}, {Lamers}, {de Mink}, {Longmore},
  {Goodwin}, \& {Gieles}}]{bastian13}
{Bastian}, N., {Lamers}, H.~J.~G.~L.~M., {de Mink}, S.~E., {Longmore}, S.~N.,
  {Goodwin}, S.~P., \& {Gieles}, M. 2013, \mnras, 436, 2398

\bibitem[{{Bekki} \& {Freeman}(2003)}]{bekki03}
{Bekki}, K. \& {Freeman}, K.~C. 2003, \mnras, 346, L11

\bibitem[{{Bekki} \& {Yong}(2012)}]{bekki12}
{Bekki}, K. \& {Yong}, D. 2012, \mnras, 419, 2063

\bibitem[{{Blackwell} {et~al.}(1986){Blackwell}, {Booth}, {Haddock}, {Petford},
  \& {Leggett}}]{blackwell86fea}
{Blackwell}, D.~E., {Booth}, A.~J., {Haddock}, D.~J., {Petford}, A.~D., \&
  {Leggett}, S.~K. 1986, \mnras, 220, 549

\bibitem[{{Blackwell} {et~al.}(1979{\natexlab{a}}){Blackwell}, {Ibbetson},
  {Petford}, \& {Shallis}}]{blackwell79feb}
{Blackwell}, D.~E., {Ibbetson}, P.~A., {Petford}, A.~D., \& {Shallis}, M.~J.
  1979{\natexlab{a}}, \mnras, 186, 633

\bibitem[{{Blackwell} {et~al.}(1995){Blackwell}, {Lynas-Gray}, \&
  {Smith}}]{blackwell95fea}
{Blackwell}, D.~E., {Lynas-Gray}, A.~E., \& {Smith}, G. 1995, \aap, 296, 217

\bibitem[{{Blackwell} {et~al.}(1979{\natexlab{b}}){Blackwell}, {Petford}, \&
  {Shallis}}]{blackwell79fea}
{Blackwell}, D.~E., {Petford}, A.~D., \& {Shallis}, M.~J. 1979{\natexlab{b}},
  \mnras, 186, 657

\bibitem[{{Blackwell} {et~al.}(1980){Blackwell}, {Petford}, {Shallis}, \&
  {Simmons}}]{blackwell80fea}
{Blackwell}, D.~E., {Petford}, A.~D., {Shallis}, M.~J., \& {Simmons}, G.~J.
  1980, \mnras, 191, 445

\bibitem[{{Brown} {et~al.}(1991){Brown}, {Wallerstein}, \& {Oke}}]{brown91}
{Brown}, J.~A., {Wallerstein}, G., \& {Oke}, J.~B. 1991, \aj, 101, 1693

\bibitem[{{Buonanno} {et~al.}(1986){Buonanno}, {Caloi}, {Castellani}, {Corsi},
  {Fusi Pecci}, \& {Gratton}}]{buonanno86}
{Buonanno}, R., {Caloi}, V., {Castellani}, V., {Corsi}, C., {Fusi Pecci}, F.,
  \& {Gratton}, R. 1986, \aaps, 66, 79

\bibitem[{{Campbell} {et~al.}(2013){Campbell}, {D'Orazi}, {Yong},
  {Constantino}, {Lattanzio}, {Stancliffe}, {Angelou}, {Wylie-de Boer}, \&
  {Grundahl}}]{campbell13}
{Campbell}, S.~W., {D'Orazi}, V., {Yong}, D., {Constantino}, T.~N.,
  {Lattanzio}, J.~C., {Stancliffe}, R.~J., {Angelou}, G.~C., {Wylie-de Boer},
  E.~C., \& {Grundahl}, F. 2013, \nat, 498, 198

\bibitem[{{Campbell} {et~al.}(2012){Campbell}, {Yong}, {Wylie-de Boer},
  {Stancliffe}, {Lattanzio}, {Angelou}, {D'Orazi}, {Martell}, {Grundahl}, \&
  {Sneden}}]{campbell12}
{Campbell}, S.~W., {Yong}, D., {Wylie-de Boer}, E.~C., {Stancliffe}, R.~J.,
  {Lattanzio}, J.~C., {Angelou}, G.~C., {D'Orazi}, V., {Martell}, S.~L.,
  {Grundahl}, F., \& {Sneden}, C. 2012, \apjl, 761, L2

\bibitem[{{Carpenter}(2001)}]{carpenter01}
{Carpenter}, J.~M. 2001, \aj, 121, 2851

\bibitem[{{Carretta} {et~al.}(2009){Carretta}, {Bragaglia}, {Gratton},
  {D'Orazi}, \& {Lucatello}}]{Ca09}
{Carretta}, E., {Bragaglia}, A., {Gratton}, R., {D'Orazi}, V., \& {Lucatello},
  S. 2009, \aap, 508, 695

\bibitem[{{Carretta} {et~al.}(2010{\natexlab{a}}){Carretta}, {Bragaglia},
  {Gratton}, {Lucatello}, {Bellazzini}, {Catanzaro}, {Leone}, {Momany},
  {Piotto}, \& {D'Orazi}}]{carretta10}
{Carretta}, E., {Bragaglia}, A., {Gratton}, R.~G., {Lucatello}, S.,
  {Bellazzini}, M., {Catanzaro}, G., {Leone}, F., {Momany}, Y., {Piotto}, G.,
  \& {D'Orazi}, V. 2010{\natexlab{a}}, \apjl, 714, L7

\bibitem[{{Carretta} {et~al.}(2012){Carretta}, {D'Orazi}, {Gratton}, \&
  {Lucatello}}]{carretta12}
{Carretta}, E., {D'Orazi}, V., {Gratton}, R.~G., \& {Lucatello}, S. 2012, \aap,
  543, A117

\bibitem[{{Carretta} {et~al.}(2014){Carretta}, {D'Orazi}, {Gratton}, \&
  {Lucatello}}]{carretta141851}
---. 2014, \aap, 563, A32

\bibitem[{{Carretta} {et~al.}(2005){Carretta}, {Gratton}, {Lucatello},
  {Bragaglia}, \& {Bonifacio}}]{carretta05}
{Carretta}, E., {Gratton}, R.~G., {Lucatello}, S., {Bragaglia}, A., \&
  {Bonifacio}, P. 2005, \aap, 433, 597

\bibitem[{{Carretta} {et~al.}(2010{\natexlab{b}}){Carretta}, {Gratton},
  {Lucatello}, {Bragaglia}, {Catanzaro}, {Leone}, {Momany}, {D'Orazi},
  {Cassisi}, {D'Antona}, \& {Ortolani}}]{carretta101851}
{Carretta}, E., {Gratton}, R.~G., {Lucatello}, S., {Bragaglia}, A.,
  {Catanzaro}, G., {Leone}, F., {Momany}, Y., {D'Orazi}, V., {Cassisi}, S.,
  {D'Antona}, F., \& {Ortolani}, S. 2010{\natexlab{b}}, \apjl, 722, L1

\bibitem[{{Carretta} {et~al.}(2011){Carretta}, {Lucatello}, {Gratton},
  {Bragaglia}, \& {D'Orazi}}]{carretta11}
{Carretta}, E., {Lucatello}, S., {Gratton}, R.~G., {Bragaglia}, A., \&
  {D'Orazi}, V. 2011, \aap, 533, A69

\bibitem[{{Casetti-Dinescu} {et~al.}(2013){Casetti-Dinescu}, {Girard},
  {J{\'{\i}}lkov{\'a}}, {van Altena}, {Podest{\'a}}, \&
  {L{\'o}pez}}]{casetti-dinescu13}
{Casetti-Dinescu}, D.~I., {Girard}, T.~M., {J{\'{\i}}lkov{\'a}}, L., {van
  Altena}, W.~F., {Podest{\'a}}, F., \& {L{\'o}pez}, C.~E. 2013, \aj, 146, 33

\bibitem[{{Cassisi} {et~al.}(2008){Cassisi}, {Salaris}, {Pietrinferni},
  {Piotto}, {Milone}, {Bedin}, \& {Anderson}}]{cassisi08}
{Cassisi}, S., {Salaris}, M., {Pietrinferni}, A., {Piotto}, G., {Milone},
  A.~P., {Bedin}, L.~R., \& {Anderson}, J. 2008, \apjl, 672, L115

\bibitem[{{Castelli} \& {Kurucz}(2003)}]{castelli03}
{Castelli}, F. \& {Kurucz}, R.~L. 2003, in IAU Symp. 210, Modelling of Stellar
  Atmospheres, ed.\ N.\ Piskunov, W.\ W.\ Weiss, \& D.\ F.\ Gray (San
  Francisco, CA: ASP), A20

\bibitem[{{Cohen}(1981)}]{cohen81}
{Cohen}, J.~G. 1981, \apj, 247, 869

\bibitem[{{Cohen} \& {Mel{\'e}ndez}(2005)}]{cohen05}
{Cohen}, J.~G. \& {Mel{\'e}ndez}, J. 2005, \aj, 129, 303

\bibitem[{{Da Costa} {et~al.}(2014){Da Costa}, {Held}, \&
  {Saviane}}]{dacosta14}
{Da Costa}, G.~S., {Held}, E.~V., \& {Saviane}, I. 2014, \mnras, 438, 3507

\bibitem[{{D'Antona} {et~al.}(2009){D'Antona}, {Stetson}, {Ventura}, {Milone},
  {Piotto}, \& {Caloi}}]{dantona09}
{D'Antona}, F., {Stetson}, P.~B., {Ventura}, P., {Milone}, A.~P., {Piotto}, G.,
  \& {Caloi}, V. 2009, \mnras, 399, L151

\bibitem[{{de Mink} {et~al.}(2009){de Mink}, {Pols}, {Langer}, \&
  {Izzard}}]{demink09}
{de Mink}, S.~E., {Pols}, O.~R., {Langer}, N., \& {Izzard}, R.~G. 2009, \aap,
  507, L1

\bibitem[{{Decressin} {et~al.}(2007{\natexlab{a}}){Decressin}, {Charbonnel}, \&
  {Meynet}}]{decressin07b}
{Decressin}, T., {Charbonnel}, C., \& {Meynet}, G. 2007{\natexlab{a}}, \aap,
  475, 859

\bibitem[{{Decressin} {et~al.}(2007{\natexlab{b}}){Decressin}, {Meynet},
  {Charbonnel}, {Prantzos}, \& {Ekstr{\"o}m}}]{decressin07a}
{Decressin}, T., {Meynet}, G., {Charbonnel}, C., {Prantzos}, N., \&
  {Ekstr{\"o}m}, S. 2007{\natexlab{b}}, \aap, 464, 1029

\bibitem[{{Denisenkov} \& {Denisenkova}(1990)}]{denisenkov90}
{Denisenkov}, P.~A. \& {Denisenkova}, S.~N. 1990, Soviet Astronomy Letters, 16,
  275

\bibitem[{{Denissenkov} \& {Hartwick}(2014)}]{denissenkov14}
{Denissenkov}, P.~A. \& {Hartwick}, F.~D.~A. 2014, \mnras, 437, L21

\bibitem[{{Dickens} {et~al.}(1991){Dickens}, {Croke}, {Cannon}, \&
  {Bell}}]{dickens91}
{Dickens}, R.~J., {Croke}, B.~F.~W., {Cannon}, R.~D., \& {Bell}, R.~A. 1991,
  \nat, 351, 212

\bibitem[{{Dinescu} {et~al.}(1997){Dinescu}, {Girard}, {van Altena}, {Mendez},
  \& {Lopez}}]{dinescu97}
{Dinescu}, D.~I., {Girard}, T.~M., {van Altena}, W.~F., {Mendez}, R.~A., \&
  {Lopez}, C.~E. 1997, \aj, 114, 1014

\bibitem[{{Dotter}(2008)}]{dotter08}
{Dotter}, A. 2008, \apjl, 687, L21

\bibitem[{{Fenner} {et~al.}(2004){Fenner}, {Campbell}, {Karakas}, {Lattanzio},
  \& {Gibson}}]{fenner04}
{Fenner}, Y., {Campbell}, S., {Karakas}, A.~I., {Lattanzio}, J.~C., \&
  {Gibson}, B.~K. 2004, \mnras, 353, 789

\bibitem[{{Ferraro} {et~al.}(2009){Ferraro}, {Dalessandro}, {Mucciarelli},
  {Beccari}, {Rich}, {Origlia}, {Lanzoni}, {Rood}, {Valenti}, {Bellazzini},
  {Ransom}, \& {Cocozza}}]{ferraro09}
{Ferraro}, F.~R., {Dalessandro}, E., {Mucciarelli}, A., {Beccari}, G., {Rich},
  R.~M., {Origlia}, L., {Lanzoni}, B., {Rood}, R.~T., {Valenti}, E.,
  {Bellazzini}, M., {Ransom}, S.~M., \& {Cocozza}, G. 2009, \nat, 462, 483

\bibitem[{{Freeman}(1993)}]{freeman93}
{Freeman}, K.~C. 1993, in Astronomical Society of the Pacific Conference
  Series, Vol.~48, The Globular Cluster-Galaxy Connection, ed. G.~H. {Smith} \&
  J.~P. {Brodie}, 608

\bibitem[{{Freeman} \& {Rodgers}(1975)}]{freeman75}
{Freeman}, K.~C. \& {Rodgers}, A.~W. 1975, \apjl, 201, L71

\bibitem[{{Gratton} {et~al.}(2004){Gratton}, {Sneden}, \&
  {Carretta}}]{gratton04}
{Gratton}, R., {Sneden}, C., \& {Carretta}, E. 2004, \araa, 42, 385

\bibitem[{{Gratton} {et~al.}(2012{\natexlab{a}}){Gratton}, {Carretta}, \&
  {Bragaglia}}]{gratton12}
{Gratton}, R.~G., {Carretta}, E., \& {Bragaglia}, A. 2012{\natexlab{a}}, \aapr,
  20, 50

\bibitem[{{Gratton} {et~al.}(2010){Gratton}, {Carretta}, {Bragaglia},
  {Lucatello}, \& {D'Orazi}}]{gratton10}
{Gratton}, R.~G., {Carretta}, E., {Bragaglia}, A., {Lucatello}, S., \&
  {D'Orazi}, V. 2010, \aap, 517, A81

\bibitem[{{Gratton} {et~al.}(2003){Gratton}, {Carretta}, {Claudi}, {Lucatello},
  \& {Barbieri}}]{gratton03}
{Gratton}, R.~G., {Carretta}, E., {Claudi}, R., {Lucatello}, S., \& {Barbieri},
  M. 2003, \aap, 404, 187

\bibitem[{{Gratton} {et~al.}(2012{\natexlab{b}}){Gratton}, {Lucatello},
  {Carretta}, {Bragaglia}, {D'Orazi}, {Al Momany}, {Sollima}, {Salaris}, \&
  {Cassisi}}]{gratton121851b}
{Gratton}, R.~G., {Lucatello}, S., {Carretta}, E., {Bragaglia}, A., {D'Orazi},
  V., {Al Momany}, Y., {Sollima}, A., {Salaris}, M., \& {Cassisi}, S.
  2012{\natexlab{b}}, \aap, 539, A19

\bibitem[{{Gratton} {et~al.}(2012{\natexlab{c}}){Gratton}, {Villanova},
  {Lucatello}, {Sollima}, {Geisler}, {Carretta}, {Cassisi}, \&
  {Bragaglia}}]{gratton121851}
{Gratton}, R.~G., {Villanova}, S., {Lucatello}, S., {Sollima}, A., {Geisler},
  D., {Carretta}, E., {Cassisi}, S., \& {Bragaglia}, A. 2012{\natexlab{c}},
  \aap, 544, A12

\bibitem[{{Grundahl} {et~al.}(2002){Grundahl}, {Briley}, {Nissen}, \&
  {Feltzing}}]{grundahl02}
{Grundahl}, F., {Briley}, M., {Nissen}, P.~E., \& {Feltzing}, S. 2002, \aap,
  385, L14

\bibitem[{{Grundahl} {et~al.}(1999){Grundahl}, {Catelan}, {Landsman},
  {Stetson}, \& {Andersen}}]{grundahl99}
{Grundahl}, F., {Catelan}, M., {Landsman}, W.~B., {Stetson}, P.~B., \&
  {Andersen}, M.~I. 1999, \apj, 524, 242

\bibitem[{{Han} {et~al.}(2009){Han}, {Lee}, {Joo}, {Sohn}, {Yoon}, {Kim}, \&
  {Lee}}]{han09}
{Han}, S.-I., {Lee}, Y.-W., {Joo}, S.-J., {Sohn}, S.~T., {Yoon}, S.-J., {Kim},
  H.-S., \& {Lee}, J.-W. 2009, \apjl, 707, L190

\bibitem[{{Harris}(1996)}]{harris96}
{Harris}, W.~E. 1996, \aj, 112, 1487

\bibitem[{{Ivans} {et~al.}(1999){Ivans}, {Sneden}, {Kraft}, {Suntzeff},
  {Smith}, {Langer}, \& {Fulbright}}]{ivans99}
{Ivans}, I.~I., {Sneden}, C., {Kraft}, R.~P., {Suntzeff}, N.~B., {Smith},
  V.~V., {Langer}, G.~E., \& {Fulbright}, J.~P. 1999, \aj, 118, 1273

\bibitem[{{Johnson} \& {Pilachowski}(2010)}]{johnson10}
{Johnson}, C.~I. \& {Pilachowski}, C.~A. 2010, \apj, 722, 1373

\bibitem[{{Joo} \& {Lee}(2013)}]{joo13}
{Joo}, S.-J. \& {Lee}, Y.-W. 2013, \apj, 762, 36

\bibitem[{{Kaluzny} {et~al.}(2013){Kaluzny}, {Thompson}, {Rozyczka}, {Dotter},
  {Krzeminski}, {Pych}, {Rucinski}, {Burley}, \& {Shectman}}]{kaluzny13}
{Kaluzny}, J., {Thompson}, I.~B., {Rozyczka}, M., {Dotter}, A., {Krzeminski},
  W., {Pych}, W., {Rucinski}, S.~M., {Burley}, G.~S., \& {Shectman}, S.~A.
  2013, \aj, 145, 43

\bibitem[{{Karakas} {et~al.}(2006){Karakas}, {Fenner}, {Sills}, {Campbell}, \&
  {Lattanzio}}]{karakas06}
{Karakas}, A.~I., {Fenner}, Y., {Sills}, A., {Campbell}, S.~W., \& {Lattanzio},
  J.~C. 2006, \apj, 652, 1240

\bibitem[{{Karakas} \& {Lattanzio}(2014)}]{karakas14}
{Karakas}, A.~I. \& {Lattanzio}, J.~C. 2014, PASA in press (arXiv:1405.0062)

\bibitem[{{Kraft}(1994)}]{kraft94}
{Kraft}, R.~P. 1994, \pasp, 106, 553

\bibitem[{{Langer} {et~al.}(1993){Langer}, {Hoffman}, \& {Sneden}}]{langer93}
{Langer}, G.~E., {Hoffman}, R., \& {Sneden}, C. 1993, \pasp, 105, 301

\bibitem[{{Lardo} {et~al.}(2012){Lardo}, {Milone}, {Marino}, {Mucciarelli},
  {Pancino}, {Zoccali}, {Rejkuba}, {Carrera}, \& {Gonzalez}}]{lardo121851}
{Lardo}, C., {Milone}, A.~P., {Marino}, A.~F., {Mucciarelli}, A., {Pancino},
  E., {Zoccali}, M., {Rejkuba}, M., {Carrera}, R., \& {Gonzalez}, O. 2012,
  \aap, 541, A141

\bibitem[{{Lee} {et~al.}(2009){Lee}, {Lee}, {Kang}, {Lee}, {Han}, {Joo}, {Rey},
  \& {Yong}}]{lee09}
{Lee}, J.-W., {Lee}, J., {Kang}, Y.-W., {Lee}, Y.-W., {Han}, S.-I., {Joo},
  S.-J., {Rey}, S.-C., \& {Yong}, D. 2009, \apjl, 695, L78

\bibitem[{{Lind} {et~al.}(2011){Lind}, {Charbonnel}, {Decressin}, {Primas},
  {Grundahl}, \& {Asplund}}]{lind11}
{Lind}, K., {Charbonnel}, C., {Decressin}, T., {Primas}, F., {Grundahl}, F., \&
  {Asplund}, M. 2011, \aap, 527, A148

\bibitem[{{Marcolini} {et~al.}(2009){Marcolini}, {Gibson}, {Karakas}, \&
  {S{\'a}nchez-Bl{\'a}zquez}}]{marcolini09}
{Marcolini}, A., {Gibson}, B.~K., {Karakas}, A.~I., \&
  {S{\'a}nchez-Bl{\'a}zquez}, P. 2009, \mnras, 395, 719

\bibitem[{{Marino} {et~al.}(2012){Marino}, {Milone}, {Piotto}, {Cassisi},
  {D'Antona}, {Anderson}, {Aparicio}, {Bedin}, {Renzini}, \&
  {Villanova}}]{marino12b}
{Marino}, A.~F., {Milone}, A.~P., {Piotto}, G., {Cassisi}, S., {D'Antona}, F.,
  {Anderson}, J., {Aparicio}, A., {Bedin}, L.~R., {Renzini}, A., \&
  {Villanova}, S. 2012, \apj, 746, 14

\bibitem[{{Marino} {et~al.}(2009){Marino}, {Milone}, {Piotto}, {Villanova},
  {Bedin}, {Bellini}, \& {Renzini}}]{marino09}
{Marino}, A.~F., {Milone}, A.~P., {Piotto}, G., {Villanova}, S., {Bedin},
  L.~R., {Bellini}, A., \& {Renzini}, A. 2009, \aap, 505, 1099

\bibitem[{{Marino} {et~al.}(2014){Marino}, {Milone}, {Yong}, {Dotter}, {Da
  Costa}, {Asplund}, {Jerjen}, {Mackey}, {Norris}, {Cassisi}, {Sbordone},
  {Stetson}, {Weiss}, {Aparicio}, {Bedin}, {Lind}, {Monelli}, {Piotto},
  {Angeloni}, \& {Buonanno}}]{marino14}
{Marino}, A.~F., {Milone}, A.~P., {Yong}, D., {Dotter}, A., {Da Costa}, G.,
  {Asplund}, M., {Jerjen}, H., {Mackey}, D., {Norris}, J., {Cassisi}, S.,
  {Sbordone}, L., {Stetson}, P.~B., {Weiss}, A., {Aparicio}, A., {Bedin},
  L.~R., {Lind}, K., {Monelli}, M., {Piotto}, G., {Angeloni}, R., \&
  {Buonanno}, R. 2014, \mnras, 442, 3044

\bibitem[{{Marino} {et~al.}(2011){Marino}, {Sneden}, {Kraft}, {Wallerstein},
  {Norris}, {da Costa}, {Milone}, {Ivans}, {Gonzalez}, {Fulbright}, {Hilker},
  {Piotto}, {Zoccali}, \& {Stetson}}]{marino11}
{Marino}, A.~F., {Sneden}, C., {Kraft}, R.~P., {Wallerstein}, G., {Norris},
  J.~E., {da Costa}, G., {Milone}, A.~P., {Ivans}, I.~I., {Gonzalez}, G.,
  {Fulbright}, J.~P., {Hilker}, M., {Piotto}, G., {Zoccali}, M., \& {Stetson},
  P.~B. 2011, \aap, 532, A8

\bibitem[{{Milone} {et~al.}(2008){Milone}, {Bedin}, {Piotto}, {Anderson},
  {King}, {Sarajedini}, {Dotter}, {Chaboyer}, {Mar{\'{\i}}n-Franch},
  {Majewski}, {Aparicio}, {Hempel}, {Paust}, {Reid}, {Rosenberg}, \&
  {Siegel}}]{milone08}
{Milone}, A.~P., {Bedin}, L.~R., {Piotto}, G., {Anderson}, J., {King}, I.~R.,
  {Sarajedini}, A., {Dotter}, A., {Chaboyer}, B., {Mar{\'{\i}}n-Franch}, A.,
  {Majewski}, S., {Aparicio}, A., {Hempel}, M., {Paust}, N.~E.~Q., {Reid},
  I.~N., {Rosenberg}, A., \& {Siegel}, M. 2008, \apj, 673, 241

\bibitem[{{Milone} {et~al.}(2013){Milone}, {Marino}, {Piotto}, {Bedin},
  {Anderson}, {Aparicio}, {Bellini}, {Cassisi}, {D'Antona}, {Grundahl},
  {Monelli}, \& {Yong}}]{milone13}
{Milone}, A.~P., {Marino}, A.~F., {Piotto}, G., {Bedin}, L.~R., {Anderson}, J.,
  {Aparicio}, A., {Bellini}, A., {Cassisi}, S., {D'Antona}, F., {Grundahl}, F.,
  {Monelli}, M., \& {Yong}, D. 2013, \apj, 767, 120

\bibitem[{{Milone} {et~al.}(2009){Milone}, {Stetson}, {Piotto}, {Bedin},
  {Anderson}, {Cassisi}, \& {Salaris}}]{milone09}
{Milone}, A.~P., {Stetson}, P.~B., {Piotto}, G., {Bedin}, L.~R., {Anderson},
  J., {Cassisi}, S., \& {Salaris}, M. 2009, \aap, 503, 755

\bibitem[{{Mu{\~n}oz} {et~al.}(2013){Mu{\~n}oz}, {Geisler}, \&
  {Villanova}}]{munoz13}
{Mu{\~n}oz}, C., {Geisler}, D., \& {Villanova}, S. 2013, \mnras, 433, 2006

\bibitem[{{Nataf} {et~al.}(2013){Nataf}, {Gould}, {Pinsonneault}, \&
  {Udalski}}]{nataf13}
{Nataf}, D.~M., {Gould}, A.~P., {Pinsonneault}, M.~H., \& {Udalski}, A. 2013,
  \apj, 766, 77

\bibitem[{{Norris} \& {Da Costa}(1995)}]{norris95}
{Norris}, J.~E. \& {Da Costa}, G.~S. 1995, \apj, 447, 680

\bibitem[{{Olszewski} {et~al.}(2009){Olszewski}, {Saha}, {Knezek},
  {Subramaniam}, {de Boer}, \& {Seitzer}}]{olszewski09}
{Olszewski}, E.~W., {Saha}, A., {Knezek}, P., {Subramaniam}, A., {de Boer}, T.,
  \& {Seitzer}, P. 2009, \aj, 138, 1570

\bibitem[{{Origlia} {et~al.}(2013){Origlia}, {Massari}, {Rich}, {Mucciarelli},
  {Ferraro}, {Dalessandro}, \& {Lanzoni}}]{origlia13}
{Origlia}, L., {Massari}, D., {Rich}, R.~M., {Mucciarelli}, A., {Ferraro},
  F.~R., {Dalessandro}, E., \& {Lanzoni}, B. 2013, \apjl, 779, L5

\bibitem[{{Pasquini} {et~al.}(2002){Pasquini}, {Avila}, {Blecha}, {Cacciari},
  {Cayatte}, {Colless}, {Damiani}, {de Propris}, {Dekker}, {di Marcantonio},
  {Farrell}, {Gillingham}, {Guinouard}, {Hammer}, {Kaufer}, {Hill}, {Marteaud},
  {Modigliani}, {Mulas}, {North}, {Popovic}, {Rossetti}, {Royer}, {Santin},
  {Schmutzer}, {Simond}, {Vola}, {Waller}, \& {Zoccali}}]{pasquini02}
{Pasquini}, L., {Avila}, G., {Blecha}, A., {Cacciari}, C., {Cayatte}, V.,
  {Colless}, M., {Damiani}, F., {de Propris}, R., {Dekker}, H., {di
  Marcantonio}, P., {Farrell}, T., {Gillingham}, P., {Guinouard}, I., {Hammer},
  F., {Kaufer}, A., {Hill}, V., {Marteaud}, M., {Modigliani}, A., {Mulas}, G.,
  {North}, P., {Popovic}, D., {Rossetti}, E., {Royer}, F., {Santin}, P.,
  {Schmutzer}, R., {Simond}, G., {Vola}, P., {Waller}, L., \& {Zoccali}, M.
  2002, The Messenger, 110, 1

\bibitem[{{Prantzos} {et~al.}(2007){Prantzos}, {Charbonnel}, \&
  {Iliadis}}]{prantzos07}
{Prantzos}, N., {Charbonnel}, C., \& {Iliadis}, C. 2007, \aap, 470, 179

\bibitem[{{Ram{\'{\i}}rez} \& {Mel{\'e}ndez}(2005)}]{ramirez05}
{Ram{\'{\i}}rez}, I. \& {Mel{\'e}ndez}, J. 2005, \apj, 626, 465

\bibitem[{{Reddy} {et~al.}(2002){Reddy}, {Lambert}, {Gonzalez}, \&
  {Yong}}]{reddy02}
{Reddy}, B.~E., {Lambert}, D.~L., {Gonzalez}, G., \& {Yong}, D. 2002, \apj,
  564, 482

\bibitem[{{Renzini}(2013)}]{renzini13}
{Renzini}, A. 2013, \memsai, 84, 162

\bibitem[{{Roederer} {et~al.}(2011){Roederer}, {Marino}, \&
  {Sneden}}]{roederer11}
{Roederer}, I.~U., {Marino}, A.~F., \& {Sneden}, C. 2011, \apj, 742, 37

\bibitem[{{Rood} \& {Crocker}(1985)}]{rood85}
{Rood}, R.~T. \& {Crocker}, D.~A. 1985, in European Southern Observatory
  Conference and Workshop Proceedings, Vol.~21, European Southern Observatory
  Conference and Workshop Proceedings, ed. I.~J. {Danziger}, F.~{Matteucci}, \&
  K.~{Kjar}, 61

\bibitem[{{Scarpa} {et~al.}(2011){Scarpa}, {Marconi}, {Carraro}, {Falomo}, \&
  {Villanova}}]{scarpa11}
{Scarpa}, R., {Marconi}, G., {Carraro}, G., {Falomo}, R., \& {Villanova}, S.
  2011, \aap, 525, A148

\bibitem[{{Simmerer} {et~al.}(2013){Simmerer}, {Ivans}, {Filler}, {Francois},
  {Charbonnel}, {Monier}, \& {James}}]{simmerer13}
{Simmerer}, J., {Ivans}, I.~I., {Filler}, D., {Francois}, P., {Charbonnel}, C.,
  {Monier}, R., \& {James}, G. 2013, \apjl, 764, L7

\bibitem[{{Skrutskie} {et~al.}(2006){Skrutskie}, {Cutri}, {Stiening},
  {Weinberg}, {Schneider}, {Carpenter}, {Beichman}, {Capps}, {Chester},
  {Elias}, {Huchra}, {Liebert}, {Lonsdale}, {Monet}, {Price}, {Seitzer},
  {Jarrett}, {Kirkpatrick}, {Gizis}, {Howard}, {Evans}, {Fowler}, {Fullmer},
  {Hurt}, {Light}, {Kopan}, {Marsh}, {McCallon}, {Tam}, {Van Dyk}, \&
  {Wheelock}}]{2mass}
{Skrutskie}, M.~F., {Cutri}, R.~M., {Stiening}, R., {Weinberg}, M.~D.,
  {Schneider}, S., {Carpenter}, J.~M., {Beichman}, C., {Capps}, R., {Chester},
  T., {Elias}, J., {Huchra}, J., {Liebert}, J., {Lonsdale}, C., {Monet}, D.~G.,
  {Price}, S., {Seitzer}, P., {Jarrett}, T., {Kirkpatrick}, J.~D., {Gizis},
  J.~E., {Howard}, E., {Evans}, T., {Fowler}, J., {Fullmer}, L., {Hurt}, R.,
  {Light}, R., {Kopan}, E.~L., {Marsh}, K.~A., {McCallon}, H.~L., {Tam}, R.,
  {Van Dyk}, S., \& {Wheelock}, S. 2006, \aj, 131, 1163

\bibitem[{{Smith}(1987)}]{smith87}
{Smith}, G.~H. 1987, \pasp, 99, 67

\bibitem[{{Smith} {et~al.}(2005){Smith}, {Cunha}, {Ivans}, {Lattanzio},
  {Campbell}, \& {Hinkle}}]{smith05}
{Smith}, V.~V., {Cunha}, K., {Ivans}, I.~I., {Lattanzio}, J.~C., {Campbell},
  S., \& {Hinkle}, K.~H. 2005, \apj, 633, 392

\bibitem[{{Smith} {et~al.}(2000){Smith}, {Suntzeff}, {Cunha}, {Gallino},
  {Busso}, {Lambert}, \& {Straniero}}]{smith00}
{Smith}, V.~V., {Suntzeff}, N.~B., {Cunha}, K., {Gallino}, R., {Busso}, M.,
  {Lambert}, D.~L., \& {Straniero}, O. 2000, \aj, 119, 1239

\bibitem[{{Sneden}(1973)}]{moog}
{Sneden}, C. 1973, \apj, 184, 839

\bibitem[{{Sobeck} {et~al.}(2011){Sobeck}, {Kraft}, {Sneden}, {Preston},
  {Cowan}, {Smith}, {Thompson}, {Shectman}, \& {Burley}}]{sobeck11}
{Sobeck}, J.~S., {Kraft}, R.~P., {Sneden}, C., {Preston}, G.~W., {Cowan},
  J.~J., {Smith}, G.~H., {Thompson}, I.~B., {Shectman}, S.~A., \& {Burley},
  G.~S. 2011, \aj, 141, 175

\bibitem[{{Spite} {et~al.}(2005){Spite}, {Cayrel}, {Plez}, {Hill}, {Spite},
  {Depagne}, {Fran{\c c}ois}, {Bonifacio}, {Barbuy}, {Beers}, {Andersen},
  {Molaro}, {Nordstr{\"o}m}, \& {Primas}}]{spite05}
{Spite}, M., {Cayrel}, R., {Plez}, B., {Hill}, V., {Spite}, F., {Depagne}, E.,
  {Fran{\c c}ois}, P., {Bonifacio}, P., {Barbuy}, B., {Beers}, T., {Andersen},
  J., {Molaro}, P., {Nordstr{\"o}m}, B., \& {Primas}, F. 2005, \aap, 430, 655

\bibitem[{{Stetson}(1981)}]{stetson81}
{Stetson}, P.~B. 1981, \aj, 86, 687

\bibitem[{{Stetson} \& {Pancino}(2008)}]{stetson08}
{Stetson}, P.~B. \& {Pancino}, E. 2008, \pasp, 120, 1332

\bibitem[{{Thompson} {et~al.}(2010){Thompson}, {Kaluzny}, {Rucinski},
  {Krzeminski}, {Pych}, {Dotter}, \& {Burley}}]{thompson10}
{Thompson}, I.~B., {Kaluzny}, J., {Rucinski}, S.~M., {Krzeminski}, W., {Pych},
  W., {Dotter}, A., \& {Burley}, G.~S. 2010, \aj, 139, 329

\bibitem[{{Ventura} {et~al.}(2009){Ventura}, {Caloi}, {D'Antona}, {Ferguson},
  {Milone}, \& {Piotto}}]{ventura09}
{Ventura}, P., {Caloi}, V., {D'Antona}, F., {Ferguson}, J., {Milone}, A., \&
  {Piotto}, G.~P. 2009, \mnras, 399, 934

\bibitem[{{Ventura} \& {D'Antona}(2005)}]{ventura05}
{Ventura}, P. \& {D'Antona}, F. 2005, \apjl, 635, L149

\bibitem[{{Villanova} {et~al.}(2010){Villanova}, {Geisler}, \&
  {Piotto}}]{villanova10}
{Villanova}, S., {Geisler}, D., \& {Piotto}, G. 2010, \apjl, 722, L18

\bibitem[{{Yong} \& {Grundahl}(2008)}]{yong081851}
{Yong}, D. \& {Grundahl}, F. 2008, \apjl, 672, L29

\bibitem[{{Yong} {et~al.}(2009){Yong}, {Grundahl}, {D'Antona}, {Karakas},
  {Lattanzio}, \& {Norris}}]{yong09}
{Yong}, D., {Grundahl}, F., {D'Antona}, F., {Karakas}, A.~I., {Lattanzio},
  J.~C., \& {Norris}, J.~E. 2009, \apjl, 695, L62

\bibitem[{{Yong} {et~al.}(2008{\natexlab{a}}){Yong}, {Grundahl}, {Johnson}, \&
  {Asplund}}]{yong08nh}
{Yong}, D., {Grundahl}, F., {Johnson}, J.~A., \& {Asplund}, M.
  2008{\natexlab{a}}, \apj, 684, 1159

\bibitem[{{Yong} {et~al.}(2003){Yong}, {Grundahl}, {Lambert}, {Nissen}, \&
  {Shetrone}}]{yong03}
{Yong}, D., {Grundahl}, F., {Lambert}, D.~L., {Nissen}, P.~E., \& {Shetrone},
  M.~D. 2003, \aap, 402, 985

\bibitem[{{Yong} {et~al.}(2005){Yong}, {Grundahl}, {Nissen}, {Jensen}, \&
  {Lambert}}]{yong05}
{Yong}, D., {Grundahl}, F., {Nissen}, P.~E., {Jensen}, H.~R., \& {Lambert},
  D.~L. 2005, \aap, 438, 875

\bibitem[{{Yong} {et~al.}(2008{\natexlab{b}}){Yong}, {Mel{\'e}ndez}, {Cunha},
  {Karakas}, {Norris}, \& {Smith}}]{yong086712}
{Yong}, D., {Mel{\'e}ndez}, J., {Cunha}, K., {Karakas}, A.~I., {Norris}, J.~E.,
  \& {Smith}, V.~V. 2008{\natexlab{b}}, \apj, 689, 1020

\bibitem[{{Yong} {et~al.}(2013){Yong}, {Mel{\'e}ndez}, {Grundahl}, {Roederer},
  {Norris}, {Milone}, {Marino}, {Coelho}, {McArthur}, {Lind}, {Collet}, \&
  {Asplund}}]{yong13}
{Yong}, D., {Mel{\'e}ndez}, J., {Grundahl}, F., {Roederer}, I.~U., {Norris},
  J.~E., {Milone}, A.~P., {Marino}, A.~F., {Coelho}, P., {McArthur}, B.~E.,
  {Lind}, K., {Collet}, R., \& {Asplund}, M. 2013, \mnras, 434, 3542

\bibitem[{{Yong} {et~al.}(2014){Yong}, {Roederer}, {Grundahl}, {Da Costa},
  {Karakas}, {Norris}, {Aoki}, {Fishlock}, {Marino}, {Milone}, \&
  {Shingles}}]{yong14}
{Yong}, D., {Roederer}, I.~U., {Grundahl}, F., {Da Costa}, G.~S., {Karakas},
  A.~I., {Norris}, J.~E., {Aoki}, W., {Fishlock}, C.~K., {Marino}, A.~F.,
  {Milone}, A.~P., \& {Shingles}, L.~J. 2014, \mnras, 441, 3396

\bibitem[{{Zoccali} {et~al.}(2009){Zoccali}, {Pancino}, {Catelan}, {Hempel},
  {Rejkuba}, \& {Carrera}}]{zoccali09}
{Zoccali}, M., {Pancino}, E., {Catelan}, M., {Hempel}, M., {Rejkuba}, M., \&
  {Carrera}, R. 2009, \apjl, 697, L22

\end{thebibliography}
\end{document}